\documentclass[a4paper,12pt]{article}

\usepackage{amssymb,amsmath,latexsym,color,stmaryrd,graphicx,bm}

\usepackage{comment}

\usepackage{appendix} 

\newcommand{\N}{\mathbb{N}}
\newcommand{\Z}{\mathbb{Z}}
\newcommand{\R}{\mathbb{R}}
\newcommand{\C}{\mathbb{C}}
\newcommand{\W}{{\mathcal W}}

\newcommand{\p}{\partial}

\newcommand{\e}{\varepsilon}
\newcommand{\dl}{\delta}
\newcommand{\pphi}{\varphi}

\newcommand{\Ord}{{\mathcal O}}

\newtheorem{theorem}{Theorem}[section]
\newtheorem{lemma}[theorem]{Lemma}
\newtheorem{proposition}[theorem]{Proposition}
\newtheorem{definition}[theorem]{Definition}
\newtheorem{remark}[theorem]{Remark}

\newcounter{hypo}

\newenvironment{hyp}{
 \begin{enumerate}
\setcounter{enumi}{\value{hypo}} \item}{\stepcounter{hypo} \end{enumerate}}

\makeatletter
 \@addtoreset{equation}{section}
 \makeatother
 
\title{Landau-Zener Formula in a ``Non-Adiabatic'' regime for avoided crossings}

\author{{ }\\ Takuya Watanabe\\
\small Department of Mathematical Sciences, \\
\small Ritsumeikan University, \\
\small 1-1-1 Noji Higashi, Kusatsu, Shiga,\\
\small 525-8577 Japan.\\
\scriptsize t-watana@se.ritsumei.ac.jp \and
{ }\\ Maher Zerzeri\\
 \small LAGA {\scriptsize - UMR7539 CNRS}, \\
 \small Universit\'e Sorbonne Paris Nord,\\
  \small 99, avenue J.-B. Cl\'ement,\\
  \small 93430 Villetaneuse, France.\\ 
  \scriptsize zerzeri@math.univ-paris13.fr
}

\date{}

\begin{document}

\maketitle

\bigskip

\begin{abstract}
We study a two-level transition probability for a finite number of avoided crossings with a small interaction.
Landau-Zener formula, which gives the transition probability for one avoided crossing as 
{\scriptsize $e^{-\pi\frac{\varepsilon^{2}}{h}}$}, 
implies that the parameter $h$ and the interaction  $\varepsilon$ play an opposite role when both tend to $0$. 
The exact WKB method produces a generalization of that formula under the optimal regime 
$\scriptstyle{\frac{h}{\varepsilon^2}}$ tends to~0.
In this paper, we investigate the case $\scriptstyle{\frac{\varepsilon^2}{h}}$ tends to  0, 
called ``{\it non-adiabatic}'' regime. This is done
by reducing the associated Hamiltonian to a microlocal branching model which  
gives us the asymptotic expansions  of the local transfer matrices. \\

\vskip2cm


\end{abstract}


\section{Introduction and result}\label{int0}
\setcounter{equation}{0}

In this paper, we study a first order ordinary $2\times 2$ system: 
\begin{equation}\label{e1}
ih{\frac{d}{dt}}\psi(t)=H(t;\varepsilon)\psi(t) \quad \text{for}\ t\in \R,
\end{equation}
where $h$ is a positive parameter and $H(t;\e)$ is the $2\times 2$ matrix of the form:
$$
H(t;\varepsilon):= \left(
\begin{array}{cc}
V(t) & \varepsilon\\
\varepsilon & -V(t)
\end{array}
\right)\quad \text{with}\ \e>0.
$$ 
The system \eqref{e1} is a model equation of the time-dependent Schr\"{o}dinger equation 
whose Hamiltonian $H(t;\e)$ describes a two-level system depending on a time variable $t$ 
with a vector-valued solution $\psi(t)={}^{t}\big(\psi_1(t),\psi_2(t)\big)\in \mathbb C^2$.  
The diagonal entries of $H(t;\e)$ are two non-perturbed energies $\pm V(t)$, 
where $V(t)$ is a real-valued smooth function on $\mathbb R$.
The zeros of $V(t)$ mean the crossing points of them, 
and there the eigenstates corresponding to two non-perturbed energies involve an interaction each other.
Taking into account of such an interaction as a positive parameter $\e$ in the off-diagonals of $H(t;\e)$, 
we can treat the energy-levels of the whole system as the eigenvalues of $H(t;\e)$, 
that is $\pm \sqrt{V(t)^2 + \e^2}$. 
The difference of the eigenvalues (spectral gap) is given by $2\sqrt{V(t)^2 + \e^2}$ 
and it is strictly positive for all $t\in \mathbb R$.
If the function $V$ vanishes at some points then the minimum of the gap is exactly $2\e$ and attained 
at the crossing points.\\   

In a quantum chemistry, the situation described above is called {\it avoided crossing}. 
The point $t_0\in \mathbb R$ such that $V(t_0)=0$ plays a very peculiar role.
In mathematical aspect, the eigenvalues cross each other in a complex plane and 
the crossing points $\zeta_\pm \in \mathbb C$ with $\pm {\rm Im}\, \zeta_\pm >0$ 
such that $V(\zeta_\pm)^2+\e^2 = 0$ 
are called {\it turning points}, which also play an important role within some well-studied limits.\\ 

In the classical mechanics, the eigenstates corresponding to the eigenvalues 
$\pm \sqrt{V(t)^2+\e^2}$ propagate along them and 
especially, under the existence of the strictly positive difference between two eigenvalues, 
transitions between the two eigenvalues can not happen. 
In the quantum mechanics, however, transitions between them can be found. 
It is a natural problem to study the transition probability between two eigenvalues and 
its contribution which might come from the avoided crossing.\\ 

While $\e$ stands for an ``{\it interaction}'',
$h$ is so-called adiabatic parameter 
which can be regarded as a semi-classical one in a mathematical literature. 
For $V(t) := vt, (v>0)$ the transition probability $P(\e,h)$ is given by 
$P(\e,h) = e^{-\frac{\pi}{v}\frac{\e^2}{h}}$,  so-called Landau-Zener formula, 
(see \cite{La65} and \cite{Ze32_01}). 
From this formula, $P(\e,h)$  tends to $1$ when $\varepsilon$ goes to $0$ (for a fixed $h$), whereas 
it decays exponentially when $h$ goes to $0$ (for a fixed $\e$). Thus,
the adiabatic effect ($h\to 0$) and the interaction effect 
($\varepsilon \to 0$) play the ``opposite'' roles in the asymptotic expansions 
of the transition probability.\\ 

The generalization of the Landau-Zener formula based on so-called ``{\it Adiabatic Theorem}'', 
that is, the exponential decay property of the transition probability 
in the adiabatic limit $h \to 0$ (for a fixed $\e$),  
has been investigated by many authors, see for instance \cite{HaJo05_01}. 
For a detailed background of this subject, we can  
consult the books \cite{Ha94_01}, \cite{Te03_01} and their references given therein.  
Such a problem can be studied for a more general setting 
where a Hamiltonian is a self-adjoint operator on a separable Hilbert space 
by a functional analysis as in \cite{Ka50_01}, \cite{Ne80_01} 
and by a microlocal theory as in \cite{Ma94_01}. \\

On the other hand, in the setting where a Hamiltonian 
$H(t;\e)$ is a $2\times 2$ matrix-valued operator, 
generalizations of $V(t)$ are even interesting problems 
and have been studied more concretely by means of a WKB approach. 
For example, either $V(t)$ vanishes at more than one point on $\R$ 
(several avoided crossings as in \cite{JoMiPf91_01}) 
or does to order $n$ (tangential avoided crossing as in \cite{Wa06_01}). 
Moreover, in the latter setting, we can consider the case 
where a Hamiltonian has a small eigenvalue gap ($\e \to 0$). 
However, when both parameters $\e$ and $h$ tend to $0$, a WKB approach 
encounters a difficulty caused by the confluence of turning points 
as in Remark \ref{remWKB}. 
In fact, the exact WKB method (see Appendix \ref{Rev_WKB}) points out essentially
two different regimes ${{\scriptstyle\frac{h}{\e^2}}} \to 0$ and 
${{\scriptstyle\frac{\e^2}{h}}} \to 0$ in the Wronskian formula. \\

These last regimes are implied by Landau-Zener formula and 
also indicated by \cite{CoLoPo99_01}, \cite{FeLa17} and \cite{Ro04_01}. 
More precisely, the authors in \cite{CoLoPo99_01} strongly emphasized these regimes in the case of avoided crossing on the phase space.
For a system of the time-dependent Schr\"{o}dinger equation with initial data of wave packets, 
an algorithm given in \cite{FeLa17} helps to explain the reason 
why these regimes might appear in an avoided crossing situation.  
In \cite{Ro04_01} the author investigated the asymptotics of the related transfer 
matrix between wave packets with Gaussian profiles 
as both parameters tend to $0$ by means of matching method in the overlapping region.\\

In the regime ${{\scriptstyle\frac{h}{\e^2}}} \to 0$, the exact WKB method works even well, 
so we can regard this regime as ``{\it adiabatic}'' (see Theorem \ref{thmWKB}).
Our interest is the regime $\frac{\e^2}{h}$ goes to $0$, the exact WKB method is not valid 
at crossing points. Hence we call this regime ``{\it non-adiabatic}".\\ 

As mentioned above, we focus on the asymptotic behavior of the transition probability 
in the case of a ``{\it non-adiabatic}'' regime, that is both parameters 
$\e$ and $h$ tend to $0$ with $\frac{\e^2}{h}$ goes to $0$. 
In Theorem \ref{mainthm}, we show that the asymptotic expansion 
of the transition probability
depends on the parity of the number of zeros of $V$
and we give a precise expression of
the prefactor of $\scriptstyle \pi \frac{\e^2}{h}$ (see (\ref{Cn})).
Moreover, as in Remark \ref{vanishCn} {\bf (C)}, 
we derive a ``Bohr-Sommerfeld'' type quantization rule
from the condition that the prefactor $C_n(h)$ vanishes. 
Notice that each condition in the case where $n$ is even or odd implies 
the necessary condition for a ``{\it complete reflection}'' 
(if $n$ is even) or a ``{\it complete transmission}'' (if $n$ is odd).\\

Next, we give the precise assumptions and state our results.

\begin{hyp} \label{A1}
The function $V$ is real on the real axis and analytic in the following complex domain
$$
{\mathcal S}: =\{ t\in\mathbb C\ ;\ |\textrm{Im}\,t| < (\tan\theta_0)\, \langle\textrm{Re}\,t\rangle\},
$$
for some $\theta_0 \in (0,\frac{\pi}{2})$.
\end{hyp}
Here $\langle s \rangle=(1+|s|^2)^{\frac{1}{2}}$ for $s\in \mathbb R$. 
We denote by $\textrm{Re}(t),$  $\textrm{Im}(t)$ 
the real, the imaginary part of  
$t\in \mathbb C$, respectively.

\begin{hyp} \label{A2}
There exist two non-zero real constants 
$E_r, E_l$ and $\delta >1$ such that 
$$
V(t)=\left\{
\begin{array}{ll}
E_r + {\mathcal O}\left(|t|^{-\delta}\right) 
&\ \textrm{as}\ \textrm{Re}\,t\to +\infty \ \textrm{in}\ {\mathcal S},\\
E_l+ {\mathcal O}\left(|t|^{-\delta}\right) 
&\ \textrm{as}\ \textrm{Re}\,t\to -\infty \ \textrm{in}\ {\mathcal S}.
\end{array} \right.
$$
\end{hyp}

Under the analyticity condition \ref{A1} and the asymptotic conditions at infinity \ref{A2}, 
we can define the transition probability as follows:

\begin{definition}\label{def_P}
The transition probability $P(\varepsilon, h)$ is defined by 
$$
P(\varepsilon, h):=|s_{21}(\varepsilon, h)|^2=|s_{12}(\varepsilon, h)|^2,
$$
where $s_{21}$ and $s_{12}$ are off-diagonal entries of the scattering matrix 
$S(\e, h)$ given by \eqref{def_S}.
\end{definition}

As mentioned before, the distance between the two energy-levels
is larger than $2\e$ and this minimum is attained when $V(t)$ vanishes. 
We assume the following crossing condition which describes that a  
non-degenerate avoided crossings happen finite times:

\begin{hyp} \label{A3}
The potential $V$ has a finite number of zeros $t_1>\cdots>t_n$ on $\mathbb R$ 
and the order of every zero is 1.
\end{hyp}

We set $v_k:=|V'(t_k)|$ for $k=1,\cdots,n$. 
The assumption \ref{A3} implies that $v_k \neq 0$ for all $k$. 
Without loss of generality, we may assume that $V'(t_1)>0$.  
Notice that the sign of $V(t)$ for $t<t_n$ changes depending on the parity of $n$.\\

Under the regime ${\scriptstyle\frac{\e^2}{h}}\to 0$, the non-perturbed energies 
$\pm V(t)$ is more dominant than the interaction $\e$ even in the adiabatic limit $h\to 0$, 
so that the exponential decay of the transition probability 
for avoided crossings can not be found any longer.
Our main result is the following:

\begin{theorem}[Non-adiabatic regime]\label{mainthm}
Assume \ref{A1}, \ref{A2} and \ref{A3}. 
Then there exist $\mu_0>0$ and $h_0>0$ small enough such that, 
for any $\e$ and $h$ with ${\scriptstyle\frac{\e^2}{h}}  \in (0, \mu_0]$ 
and $h \in (0,h_0]$, the transition probability 
$P(\varepsilon, h)$ has the asymptotic expansions:  
$$
P(\varepsilon, h) =\left\{
\begin{array}{llll}
1-\pi C_n(h) \frac{\varepsilon^2}{h}&+&
{\mathcal O}\left( \sqrt{h} \frac{\e^2}{h} \right) + 
{\mathcal O}\left(\left(\frac{\varepsilon^2}{h}\right)^{\frac{3}{2}} \right)
 \qquad \textrm{if} \ n \ \textrm{is odd},\\[19pt]
{\quad } \pi C_n(h) \frac{\varepsilon^2}{h}&+& 
{\mathcal O}\left(\sqrt{h} \frac{\e^2}{h} \right) + 
{\mathcal O}\left(\left(\frac{\varepsilon^2}{h}\right)^{\frac{3}{2}} \right)
 \qquad \textrm{if} \ n \ \textrm{is even},
\end{array}
\right.
$$
where $C_1(h)=\frac{1}{v_1}$ and $C_n(h)$ for $n\geq 2$ is given by 
\begin{align}\label{Cn}
C_n(h) &= \sum_{k=1}^n \frac{1}{v_k} 
+ 2\sum_{k=2}^n \sum_{j=1}^{k-1} \frac{1}{\sqrt{v_j v_k} } 
\cos \left[  \frac{2}{h} \int_{t_k}^{t_j} V(t) dt  +  
\frac{(-1)^j - (-1)^k}{2} \frac{\pi}{2} \right].
\end{align}
\end{theorem}

\begin{remark}

\begin{description}

\item[{}]
 
\item[(A) One crossing point:] When $n=1$, we recover the Landau-Zener formula.

\item[(B) Dependence on the parity:] The asymptotic expansions of $P(\varepsilon, h)$ 
as ${\scriptstyle\frac{\varepsilon^2}{h}}\to 0$ 
depending on the parity of $n$ imply that the time evolutions of the eigenstates 
propagate along the non-perturbed energies $\pm V(t)$ instead of the energies 
of whole system $\pm \sqrt{V(t)^2 + \varepsilon^2}$. 
 
\item[(C) Bohr-Sommerfeld quantization rule:]\label{vanishCn}
The prefactor $C_n(h)$ is independent of $\e$ and, in this sense, 
the expression of $C_n(h)$ is refined more than that of our previous note {\rm\cite{WaZe12_01}}. 
In particular, $C_n(h)$ for $n\geq2$ may vanish, 
while $C_1(h)$ does not. 
Taking $n=2$ and $v_1=v_2$, for example, one sees 
$$
C_2(h) = \frac{2}{v_1} 
\left( 1 + \sin \left[ \frac{2}{h}\int_{t_2}^{t_1} V(t)\, dt \right]\right).
$$

The condition that $C_2(h)$ vanishes is equivalent to 
\begin{equation}\label{BS_like}
\int_{t_2}^{t_1} V(t)\, dt = \left(N - \frac{1}{4}\right)\pi h\,,
\end{equation}
for some integer $N$. 
This may be understood as a Bohr-Sommerfeld quantization rule. 
Notice that the phase shift in \eqref{BS_like} appears in the case where the action
enclosed by two different characteristics is quantized 
as in {\rm \cite{FuMaWa19_01}}. 

\end{description}
\end{remark}

In an ``adiabatic'' regime, that is  $(\e,h) \to (0,0)$ with ${\scriptstyle \frac{h}{\e^2}} \to 0$, 
the turning points and the action integrals are crucial. 
For each $k$ $(k=1,\cdots,n)$, there exist two simple turning points 
$\zeta_k(\varepsilon)$ and $\overline{\zeta_k(\varepsilon)}$, 
that is, simple zeros in ${\mathcal S}$ of $2\sqrt{-\det H(t;\varepsilon)}$ close to $t_k$,
where $\det H(t;\varepsilon):=-V(t)^2-\varepsilon^2$. 
For $k=1,\cdots,n$, we define the action integral $A_k(\varepsilon)$ by
\begin{equation}\label{Actionk}
A_k(\varepsilon)=2\int_{t_k}^{\zeta_k(\varepsilon)}{\sqrt{V(t)^2+\varepsilon^2}}\,dt, 
\end{equation}
where each path is the segment from $t_k$ to $\zeta_k(\varepsilon)$ on $\C$ 
and the branch of each square root is $\varepsilon$ at $t=t_k$.  
On this branch, ${\rm Im}\, A_k(\varepsilon)$ is positive and of $\Ord(\e^2)$ as $\e \to 0$. 
Notice that the distance in Lemma \ref{wkblem2} can be written by ${\rm Im}\, A_k(\varepsilon)$ 
so that the ratio $\scriptstyle \frac{h}{\e^2}$ appears. 
This case was treated in \cite{Jo94_01} when $V$ vanishes 
at only one point ($n=1$ in \ref{A3}), but for more general Hamiltonians. 
We extend slightly this result in our simpler framework 
but with avoided crossings ($n\geq 2 $ in \ref{A3}), by using the exact WKB method reviewed in Appendix \ref{Rev_WKB}. 

\begin{theorem}[Adiabatic regime]\label{thmWKB}
Assume \ref{A1}, \ref{A2} and \ref{A3}.  
The transition probability as 
$(\e,h) \to (0,0)$ with ${\scriptstyle \frac{h}{\e^2}}\to 0$ is given by
\begin{equation}\label{thmWKBformula}
P(\varepsilon, h)= \left|
\sum_{k\in\mathcal{K}} (-1)^k e^{\frac{i}{h}\big( A_k(\varepsilon) + {\rm Re}\, A_k(\e)  
-\overset{k-1}{\underset{j=1}{\sum}} R_{j}(\varepsilon) \big)}
\right|^2 + {\mathcal O}\left(
\frac{h}{\varepsilon^2}\, e^{-\frac{2\alpha(\varepsilon)}{h}}
\right),
\end{equation}
where $\alpha (\varepsilon) = \underset{k\in\mathcal{K}}{\min}\, 
\big(\textnormal{Im}\,A_k(\varepsilon)\big) >0$, 
an action integral $R_{j}(\varepsilon)$ is given by
\begin{equation}\label{Actionjk}
R_{j}(\varepsilon) = 2\int_{t_{j+1}}^{t_j}
{\sqrt{V(t)^2+\varepsilon^2}}\,dt,\quad \textrm{for} \ 1\leq j\leq n -1
\end{equation}
and $\mathcal{K}$ is the set of $k\in\{1,2,\cdots,n\}$ 
which attains $\max\{v_1,\cdots,v_n\}$. 
Recall that $v_k=|V'(t_k)|>0$ for $k=1,\ldots,n$.
\end{theorem}


In the case of one avoided non-degenerate crossing (i.e., $n = 1$), 
we have only one transfer matrix (see Proposition \ref{S-represent})
and take into account the off-diagonal entry only.
Then the error in \eqref{thmWKBformula} is ${\mathcal O}(h)$ uniformly with respect to $\varepsilon$
and the formula \eqref{thmWKBformula} recovers the previous results of \cite{Jo94_01}, \cite{Ro04_01}, \cite{Wa06_01} in our setting. 
In the case of more than one avoided crossing, \eqref{thmWKBformula} also does the previous work \cite{JoMiPf91_01}.
Especially, the exponentially decaying rate of the transition probability in \eqref{thmWKBformula} 
is characterized by the maximum of the derivative of $V$ at crossing points, 
so that the global configuration of non-perturbed energy (the number of zeros) does not contribute 
to the asymptotic expansions of the transition probability in the regime ${{\scriptstyle\frac{h}{\e^2}}} \to 0$. 

The critical regime $\e \sim C \sqrt{h}$ with some strictly positive constant $C$,
as $(\varepsilon,h)\to (0,0)$, 
 was treated by G.-A.~Hagedorn (see \cite{Ha91_01}).
In that paper, the author obtained the exponential decay property of the transition probability 
under a more general setting than Landau-Zener model. The proof is 
essentially based on the properties of ``parabolic cylinder functions''. \\

\noindent
{\it Outline of the proof of main result {\rm (}Theorem \ref{mainthm}\,{\rm )}}:
As mentioned above, the exact WKB method is not valid near the crossing point in our regime, 
i.e., ${\scriptstyle\frac{\e^2}{h}}\to 0$. To avoid such a difficulty, we consider microlocal solutions 
associated to the branching model at the crossing point.
This branching model is well-known in the case of single-valued Schr\"odinger 
equation since the work of Helffer-Sj\"ostrand (see \cite{HeSj89_01}).
The case of $2\times 2$ matrix type is treated by \cite{KaRo93_01} and also \cite{FuLaNe09_01}. 
While all these works are 1-parameter problems, our situation is 2-parameters one.
In fact, near the crossing point the matrix-valued operator \eqref{eq1_0} 
is reduced to the branching model given by \eqref{eq1_1}, 
with a new semi-classical parameter $\mu:=\scriptstyle{\frac{\e^2}{h}}$ (see Subsection \ref{MicroReduction}). 
But this reduction is valid only in the disc of radius of order ${\mathcal O}(\sqrt{h})$,
centered at the crossing point. This restriction on the radius is 
due to the application of some kind of Neumann's lemma, (see Lemma \ref{M_Id}).
This reduction gives a correspondence between the solutions of our model 
and the branching one via some semi-classical
Fourier integral operator with respect to $\mu$, (see Proposition \ref{relation_model}).
Since the four solutions of the branching model are explicit (see Appendix \ref{Sol_BM}), 
then we obtain the asymptotic expansions of the corresponding ones of our original  
equation by a stationary phase method,
(see Proposition \ref{Asymlocal2}). 
These asymptotic behaviors are valid in some 
pointed interval centered at the crossing point, whose length is of order ${\mathcal O}(\sqrt{h})$.

\smallskip\noindent
On the other hand, the asymptotic expansions of exact WKB solutions can be extended outside 
 a disc of size ${\mathcal O}(\sqrt{h})$, centered at the crossing point, 
 since we control the error term (see \eqref{error_annu}).
Then, for $\lambda_0$ fixed large enough, 
the asymptotic expansions of exact WKB solutions are performed in 
${\mathcal D}_0(h):=\{t\in \mathbb C,
\ \lambda_0\sqrt{h}\leq |t-t_{\bullet}|\leq 2\lambda_0 \sqrt{h}\}$, 
(see Proposition \ref{AsymWKB}). Here $t_{\bullet}$ is the crossing point.

\smallskip\noindent
The next step is to compare the microsupports of such exact WKB solutions 
and those obtained by the microlocal reduction 
for $t$ real in ${\mathcal D}_0(h)$, as in \cite{FuLaNe09_01}, \cite{FuRa98_01}, \cite{Ra96_01}. 
This argument guaranties the one to one co-linear relation between these solutions.
Then we get the asymptotic expansion of the local transfer matrix (see Proposition \ref{T_0}).

\smallskip\noindent
In the conclusion,  we give the asymptotic behavior of the transition probability, 
since the scattering matrix is a product of local transfer matrices associated at each crossing point, 
(see Subsection \ref{SSprodT}). \\

\begin{remark} We would like to emphasize that from the technical viewpoint of the microlocal connection, 
we can classify the ``non-adiabatic" regime into two sub-regimes: 

\begin{itemize}
 
\item The first regime is $\e = {\mathcal O}(h)$.
The connection between the microlocal solutions obtained by the branching model
and the exact WKB method is carried out in an $ {\mathcal O}(1)$-neighborhood
(independent of the parameter $h$) of the crossing point
by using the strategy developed by Fujii\'e-Lasser-N\'ed\'elec in {\rm\cite{FuLaNe09_01}}.
Indeed, the microlocal connection works better than the next sub-regime.
 
\item The second one is $h \ll \e \leq o(\sqrt{h})$. The situation here is more critical.
We can connect solutions in an $ {\mathcal O}(\sqrt{h})$-annulus around the crossing point.
The microlocal argument in the annulus is performed under the specific semi-classical parameter 
$\mu:=\scriptstyle\frac{\varepsilon^2}{h}$.
This sub-regime highlights the difficulty due to two-parameter problems.
\end{itemize}

\noindent
Hence, we give the proof of our result keeping in mind the second critical sub-regime 
$\big(h\ll\varepsilon\leq o (\sqrt {h})\big)$,
which also covers the first sub-situation $\big(\varepsilon=\mathcal{O}(h)\big)$.
\end{remark}

Eventually, our result can be generalized in the case of non-globally analytic function $V(t)$. 
In this context the exact WKB  method no longer works. But, the microlocal reduction 
(see \cite[Section 1]{Sj92_01}) and the result concerning propagation of microlocal solutions
through a hyperbolic fixed point (see \cite[Proposition 1.1]{BFRZ08_01})
are very useful to get the asymptotic expansions of the scattering matrix in the ${C}^\infty$-category.  

\smallskip\noindent
{\it The rest of this paper is organized as follows.}  The next section is devoted to some preliminaries 
of the proof of Theorem \ref{mainthm}. First, we define the scattering matrix 
and reduce its computation to that of the local transfer matrix (see Subsection \ref{def_ScatteringM}).
Next, we reduce the original system to a $2\times 2$ microlocal branching model 
which is valid in an ${\mathcal O}(\sqrt{h})$-neighborhood of the crossing point (see Subsection \ref{MicroReduction}) 
and derive the asymptotic expansions of the pull-back solutions of the branching model 
based on the above microlocal reduction (see Subsection \ref{pullbacksec}). 
In Subsection \ref{wkbsec}, we show that the asymptotic expansions of the exact WKB solutions are valid 
outside of some ${\mathcal O}(\sqrt{h})$-neighborhood of the corresponding crossing point. 
And we compare the microsupports of both microlocal solutions (pull-back solutions of branching model
and exact WKB ones) in  Subsection \ref{ML connection}. 
In Section \ref{proof_thm}, we prove Theorem \ref{mainthm} by computing the asymptotic expansion 
of the local transfer matrix (see Subsection \ref{T_0}) and the product of them (see Subsection \ref{SSprodT}). 
Section \ref{proofWKBthm4} is devoted to the proof of Theorem \ref{thmWKB}.  
At last, we have placed in Appendix \ref{Rev_WKB} a short review of an exact WKB approach. 
The properties of the branching model are presented in Appendix \ref{BM_application} and 
useful lemmas are stated in Appendices \ref{Neumann} and \ref{Alg_lemmas}. \\

\smallskip\noindent 
{\it Acknowledgment.}  We are grateful to the referee
for useful remarks that have led to improvements of the exposition 
and allowed us to specify the formula \eqref{thmWKBformula} in Theorem \ref{thmWKB}.

\section{Preliminaries}\label{Pre}
\setcounter{equation}{0}
 
\subsection{{\sl Scattering matrix}}\label{def_ScatteringM}
In this subsection, let us define the scattering matrix by means of Jost solutions. 
Under the analyticity condition \ref{A1} and the asymptotic conditions at infinity \ref{A2}, 
we define four {\it Jost solutions} $J_{\pm}^r(t)$ and $J_{\pm}^l(t)$ uniquely 
which satisfy the following asymptotic conditions:
$$
\begin{aligned}
 &J_+^r(t) \sim \exp \left[+\frac{i}{h}\sqrt{E_r^2+\varepsilon^2} \, t \right]\left(
\begin{array}{c}
-\sin{\theta_r}\\
\cos{\theta_r}
\end{array}
\right)
 &&\textrm{as} \,\,\,
 \textrm{Re}\,t\to +\infty \ \ \textrm{in}\ \ {\mathcal S},\\
 &J_-^r(t) \sim \exp \left[-\frac{i}{h}\sqrt{E_r^2+\varepsilon^2} \, t \right]\left(
\begin{array}{c}
\cos{\theta_r}\\
\sin{\theta_r}
\end{array}
\right) 
 &&\textrm{as} \,\,\,
 \textrm{Re}\,t\to +\infty \ \ \textrm{in}\ \ {\mathcal S},\\
 &J_+^l(t) \sim \exp \left[+\frac{i}{h}\sqrt{E_l^2+\varepsilon^2} \, t \right]\left(
\begin{array}{c}
-\sin{\theta_l}\\
\cos{\theta_l}
\end{array}
\right)
 &&\textrm{as} \,\,\,
 \textrm{Re}\,t\to -\infty \ \ \textrm{in}\ \ {\mathcal S},\\
 &J_-^l(t) \sim \exp \left[-\frac{i}{h}\sqrt{E_l^2+\varepsilon^2} \, t \right]\left(
\begin{array}{c}
\cos{\theta_l}\\
\sin{\theta_l}
\end{array}
\right) 
 &&\textrm{as} \,\,\,
 \textrm{Re}\,t\to -\infty \ \ \textrm{in}\ \ {\mathcal S},
\end{aligned}
$$
where $\tan{2\theta_{r}}=\frac{\varepsilon}{E_{r}},$ ($0<\theta_{r} <\frac{\pi}{2}$) 
and $\tan{2\theta_{l}}=\frac{\varepsilon}{E_{l}},$ ($0<\theta_{l} <\frac{\pi}{2}$).
The pairs of Jost solutions $(J_{+}^r, J_{-}^r)$ and 
$(J_{+}^l, J_{-}^l)$ are orthonormal bases on ${\mathbb C}^2$ for any fixed $t$. 
Moreover the Jost solutions have the relations:
\begin{equation}\label{conj Jost}
J_{\pm}^r(t) = \mp \left(
\begin{array}{cc}
0 & 1\\
-1 & 0
\end{array}
\right)\overline{J_{\mp}^r(t)}, \qquad 
J_{\pm}^l(t) = \mp \left(
\begin{array}{cc}
0 & 1\\
-1 & 0
\end{array}
\right)\overline{J_{\mp}^l(t)}.
\end{equation} 
\begin{definition}
The scattering matrix $S(\varepsilon,h)$ is defined as the change of basis of Jost solutions:
\begin{equation}\label{def_S}
\left( J_+^l\  J_-^l \right)=
\left( J_+^r\  J_-^{r} \right)S(\varepsilon,h), \qquad 
S(\varepsilon,h) =\left(
\begin{array}{cc}
s_{11}(\varepsilon,h) & s_{12}(\varepsilon,h)\\
s_{21}(\varepsilon,h) & s_{22}(\varepsilon,h)
\end{array}
\right).
\end{equation}
\end{definition}

From \eqref{conj Jost}, the entries of $S(\varepsilon,h)$ satisfy
\begin{align*}
s_{11}(\e,h) = \overline{s_{22}(\e,h)} 
\quad\textrm{and}\quad s_{12}(\e,h) &= -\overline{s_{21}(\e,h)}\,.
\end{align*}
The matrix $S$ is unitary and independent of $t$.  
Hence we see that $|s_{11}(\varepsilon, h)|^2 +|s_{21}(\varepsilon, h)|^2 = 1$ and thus 
we can define the transition probability as in Definition \ref{def_P}.

Thanks to the exact WKB method recalled in Appendix \ref{Rev_WKB}, 
we obtain the following proposition which gives us the representation of the scattering matrix 
by means of the product of local transfer matrices 
between decomposed domains related to the crossing points
(see \eqref{Sbox_k} and Figure \ref{S_box}). 
This type of representation has been established in previous works
(see, for example, \cite[Section 4, identity (10)]{Fu98_01}).

\begin{figure}
\begin{center}
\scalebox{0.35}[0.35]
{\includegraphics{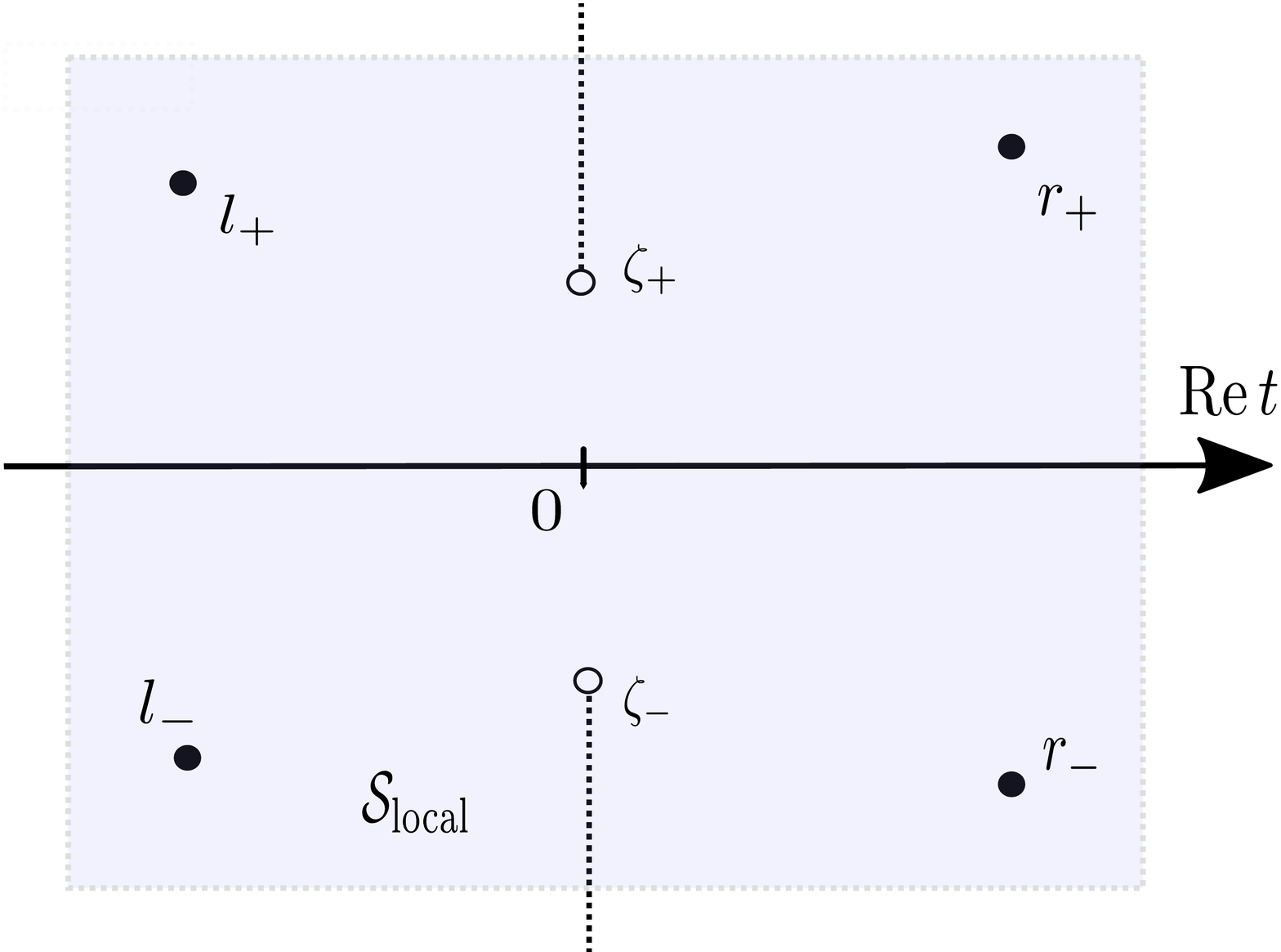}}
\caption{local setting in ${\mathcal S}_{\rm local}$}
\label{toyfig}
\end{center}
\end{figure}

\begin{proposition}\label{S-represent}
The scattering matrix ${S}(\e,h)$ is expressed by the product of the $2\times 2$ matrices 
as follows:
\begin{equation}\label{prodTra}
{S}(\e,h) = T_{r}(\e,h)^{-1}T_1(\e,h)T_{1,2}(\e,h)T_2(\e,h) \cdots 
T_{n-1,n}(\e,h)T_n(\e,h)T_{l}(\e,h),
\end{equation}
where  $\big(T_{k,k+1}(\e,h)\big)_{k=1,2,\cdots,n-1}$, $\big(T_k(\e,h)\big)_{k=1,2,\cdots,n}$ are defined in \eqref{defT_k} and      
for $\star\in \{r,l\}$, $T_\star(\e,h)$ is given by \eqref{r-inf}. 
\end{proposition}
In Appendix \ref{S_matrix_section} we prove this proposition by showing the existence of two kinds of local transfer matrices.
One of them is a change of bases with respect to the base points of the symbol function 
denoted by $T_k(\e,h)$ (see \eqref{T_kWron}), that is, it transfers from right side to left one over the crossing point $t_k$. 
The other is one with respect to the base points of the phase function denoted by $T_{k,k+1}(\e,h)$ (see \eqref{T_kk}), 
that is, it transfers from a crossing point to left one.
This is related to two situations of propagation of singularities. 

\smallskip

We conclude this subsection by claiming the following remark.
\begin{remark}\label{T_local}
Proposition \ref{S-represent} implies that,  since the local transfer matrices 
 $T_{\star}(\e,h),\  \star\in\{l,r\}$ and $T_{k,k+1}(\e,h),\,  k=1,\ldots,n-1$, can be explicitly expressed  
by actions (see \eqref{r-inf} and \eqref{T_kk}),
 it is enough to study local transfer matrices $T_k(\e,h), \, k=1,\ldots,n$ instead of 
the scattering matrix.   
Thanks to  Remark \ref{same_config}, the study of the asymptotic behaviors of $T_k(\e,h), \, k=1,\ldots,n$,
does not depend on the subscript $k$.
Then, from now on, we regard $V(t) = t + {\mathcal O}(t^2)$ in an $h$-independent neighborhood 
${\mathcal S}_{\rm local}$ of $t_k=0$. 
\end{remark}

In the rest of paper we focus on the study of the asymptotic expansion of the local transfer matrix, denoted by $T_{\rm local}$,
corresponding to the  behavior of the function $V(t)$ on some neighborhood of $t_k=0$, included in
${\mathcal S}_{\rm local}$, (see Figure \ref{toyfig}).

\subsection{{\sl Microlocal reduction near the crossing point}}\label{MicroReduction}
 
We here deal with the following form corresponding to \eqref{e1}.
\begin{equation}\label{eq1_0}
P_0 \psi(t) = \left(
\begin{matrix}
hD_t + V(t) & \e\\[7pt]
\e & hD_t - V(t)
\end{matrix}
\right)\psi(t) = 0,
\end{equation}
where $D_t = \frac{1}{i}\frac{d}{dt}$. 
Let $V(t)$ be as in the introduction. 
We can reduce the equation above \eqref{eq1_0} 
to so-called {\it branching model} of the first order $2\times 2$ system:
\begin{equation}\label{eq1_1}
Q\phi(y) = \left(
\begin{matrix}
\sqrt{2}y & \mu\\[7pt]
\mu & \sqrt{2}\mu D_y
\end{matrix}
\right)\phi(y) = 0,
\end{equation}
where $\mu = \frac{\e^2}{h}$ is a crucial small parameter in this reduction. 
This model was introduced in a single-valued case by Helffer-Sj\"ostrand \cite{HeSj89_01} (see also M\"arz \cite{Ma92_01}). 
A $2\times 2$ system depending only on one parameter was studied 
first by Kaidi-Rouleux \cite{KaRo93_01}, and also by Fermanian Kammerer-G\'erard \cite{FeGe02}, 
Colin de Verdi\`ere\cite{Co03}, Fujii\'e-Lasser-N\'ed\'elec \cite{FuLaNe09_01} in other settings. 
The properties of the solutions to the equation \eqref{eq1_1} are investigated in Appendix \ref{Sol_BM}. 

The claim of this subsection is the following reduction from \eqref{eq1_0} to \eqref{eq1_1} (see Proposition \ref{relation_model}).  
The first lemma guarantees the existence of a local smooth change of variables 
which allows us to replace $V(t)$ by a linear function near the crossing point. 
\begin{lemma}\label{local_change}
There exist a small neighborhood ${\mathcal U}$ and 
a change of variables $f : {\mathcal U} \to \tilde {\mathcal U} = f({\mathcal U})$ such that 
$f(0) = 0$, $f'(0) = 1$ and $f(t)f'(t) = V(t)$ for any $t \in {\mathcal U}$.
\end{lemma}
Note that the function $f$ is determined independently from the parameters $\e$ and $h$. 
The proof of this lemma can be done by constructing $f(t)$ concretely  for $t$ in a small neighborhood of $t= 0$ 
as follows:
\begin{equation}\label{diffeo}
f(t) = t\left( 1 + 2t \int_0^1 \!\! \int_0^1 s^2 (1-\sigma)  V''(ts\sigma) d\sigma ds \right)^{\frac{1}{2}}.
\end{equation} 

Putting $\psi_1(z) := \psi(f^{-1}(\sqrt{h}\, t))$, which consists 
of the change of variables given by the above, 
then we see that the equation \eqref{eq1_0} becomes 
\begin{equation}\label{P_1}
P_1 \psi_1(z) =
\begin{pmatrix}
D_z + z & \sqrt{\mu} (f^{-1})' (\sqrt{h}z)\\
\sqrt{\mu} (f^{-1})' (\sqrt{h}z) & D_z -z
\end{pmatrix} \psi_1(z) = 0.
\end{equation}
Recall that {\scriptsize $\mu =\frac{\e^2}{h}$}$ \to 0$ implies that $\e \to 0$ uniformly with respect to $h\in (0,h_0]$ 
for some fixed $h_0>0$. 
Notice that \eqref{P_1} is a regular perturbation problem of $\sqrt{\mu}$. 
In order to apply Lemma \ref{M_Id} in Appendix \ref{Neumann}, we should take $z$ in a bounded interval $I$ 
(i.e. $t \in \sqrt{h} I \subset {\mathcal U}$), the small parameter $\delta = \sqrt{\mu}$ 
and the $C^\infty$-function $g(z;h) = (f^{-1})'(\sqrt{h}z)$, which is bounded 
uniformly on $I\times (0, h_0]$ together with its all derivatives.
From Lemma \ref{M_Id}, there exists a $C^\infty$-matrix $M(z; \sqrt{\mu},h) = {\rm Id} + \underset{k\geq 1}{\sum}\mu^{\frac{k}{2}} M_k(z;h)$ 
such that the equation \eqref{P_1} becomes  
\begin{equation}\label{P_2}
P_2 \psi_2(z) = \begin{pmatrix}
D_z + z & \sqrt{\mu}\\
\sqrt{\mu} & D_z - z
\end{pmatrix} \psi_2(z) = 0,\qquad z\in I,
\end{equation}
where $\psi_2(z) := M(z;\sqrt{\mu}, h)\psi_1(z)$.

Now, by using a change of scaling $x = \sqrt{\mu} z$,
we can regard the equation \eqref{P_2} as a semi-classical problem with respect to $\mu$. 
\begin{equation}\label{P_3}
P_3 \psi_3(x) = \begin{pmatrix}
\mu D_x + x & \mu \\
\mu & \mu D_x -x 
\end{pmatrix}\psi_3(x) = 0,\qquad x\in \sqrt{\mu} I,
\end{equation}
where $\psi_3(x) := \psi_2(\frac{x}{\sqrt{\mu}})$.

The third lemma is so-called Egorov type theorem by means of the Fourier integral operator.  
Let $U_{\frac{\pi}{4}}$ be the Fourier integral (metaplectic) operator associated with 
the rotation $\frac{\pi}{4}$ on the phase space $T^*\R$:
$$
\kappa_{\frac{\pi}{4}} \ :\ T^*\R \ni (x,\xi) \longmapsto \frac{1}{\sqrt{2}}(x-\xi, x+\xi) \in T^*\R.
$$ 
The Fourier integral operator $U_{\frac{\pi}{4}}$ is given, in the book of Helffer-Sj\"ostrand \cite{HeSj89_01}, by 
$$
U_{\frac{\pi}{4}}[u](x) = \frac{e^{\frac{\pi}{8}i}2^{\frac{1}{4}}}{\sqrt{2\pi \mu}} 
\int_{\R} e^{\frac{i}{\mu}(-\frac{x^2}{2}+\sqrt{2}xy -\frac{y^2}{2})}u(y)\,dy 
$$
for $u$ in the space of tempered distributions.
\begin{lemma}
We denote the symbols of the diagonal entries of $P_3$ and $Q$ by
\begin{align*}
&p_{1}(x,\xi) = \xi + x, &&p_2(x,\xi) = \xi - x,\\
&q_1(y,\eta) = \sqrt{2}y, &&q_2(y,\eta) = \sqrt{2}\eta.
\end{align*}
Then the operators
$$
P_3 = \begin{pmatrix}
p_1^w(x, \mu D_x) & \mu \\ \mu & p_2^w(x, \mu D_x) 
\end{pmatrix},
\qquad  Q = \begin{pmatrix}
q_1^w(y, \mu D_y) & \mu \\ \mu & q_2^w(y, \mu D_y) 
\end{pmatrix}
$$
satisfy 
$$
U_{\frac{\pi}{4}}^{-1}  p_j^w (x, \mu D_x) U_{\frac{\pi}{4}} = ( p\circ \kappa_{\frac{\pi}{4}})^w (y, \mu D_y) = q_j(y, \mu D_y),
\quad (j=1,2).
$$
\end{lemma}

Let $\chi\in C_0^\infty(\R)$ be identically equal to $1$ near $I$. 
Then we put 
$$
\phi(y) = U_{\frac{\pi}{4}}^{-1} \left[ \chi\left( \frac{x}{\sqrt{\mu}} \right) \psi_3(x) \right](y).
$$
The equation \eqref{P_3} is equivalent to 
\begin{equation}\label{Q_1}
Q\phi(y) =  -i \sqrt{\mu} U_{\frac{\pi}{4}}^{-1} \left[\chi'\left(\frac{x}{\sqrt{\mu}}\right) \psi_3(x)\right] (y).
\end{equation}
The right-hand side of \eqref{Q_1} is of $\Ord (\mu^{\infty})$ uniformly on $\sqrt{\mu} I$.\\

Summing up, we obtain the following proposition:
\begin{proposition}\label{relation_model} 
There exist $h_0>0$ small enough and $c_0>0$ independent of $\e$ and $h$ such that 
the equation \eqref{eq1_0}  has a solution given by, as $\frac{\e^2}{h} \to 0$,    
\begin{equation}\label{phiTOpsi}
\psi(t;\e,h) = M\left( \frac{\e}{h}f(t); \frac{\e}{\sqrt{h}}, h \right) U_{\frac{\pi}{4}}\left[ \phi \right]\left(\frac{\e}{h}f(t)\right) 
 + \Ord \left(\left(\frac{\e^2}{h}\right)^{\infty}\right),
\end{equation}
uniformly on $h \in (0, h_0]$ and $t \in \{ |t| \leq c_0\sqrt{h}\}$, where $\phi$ is a solution of \eqref{Q_1}, 
and $f$ and $M$ are given respectively 
by Lemma \ref{local_change} and Lemma \ref{M_Id}.
\end{proposition}

\begin{remark}\label{scaling_regimes}
Another scaling parameter $\frac{\e}{h}$ in \eqref{phiTOpsi} implies that if $\e$ tends to $0$ faster than $h$, 
that is, $\e$ goes to $0$ quite faster than the case where $\frac{\e^2}{h} \to 0$,  
such a microlocal reduction itself with respect to $\mu$ in \S\ref{MicroReduction} works better 
like a one-parameter problem as well as {\rm \cite{FuLaNe09_01}}.
\end{remark}


\subsection{{ \sl Asymptotic expansions of the pull-back solutions of the branching model 
in some ${\mathcal{O}(\sqrt{h})}$-neighborhood of the crossing point}}
\label{pullbacksec}

In this subsection, from Proposition \ref{relation_model} and Proposition \ref{Asymlocal}, 
we derive the asymptotic behaviors of the pull-back solutions of the branching model 
on suitable intervals of order ${\mathcal{O}(\sqrt{h})}$, denoted by $I_c(h)$, that is, for some constant $c>0$, 
\begin{equation}\label{coer_interval}
I_c(h) := \{t\in \R\, ;\, c \sqrt{h} < |t| < 2c \sqrt{h}\, \}.
\end{equation}
Based on Proposition \ref{relation_model}, we denote 
by $\tilde U [\phi^\vdash](t)$ (resp. $\tilde U [\phi^\dashv](t)$, $\tilde U [\phi^\bot](t)$ and $\tilde U [\phi^\top](t)$)  
the rescaling function to 
$U_{\frac{\pi}{4}}[\phi^\vdash](x)$ (resp. $U_{\frac{\pi}{4}}[\phi^\dashv](x)$, 
$U_{\frac{\pi}{4}}[\phi^\bot](x)$ and $U_{\frac{\pi}{4}}[\phi^\top](x)$)
in Proposition \ref{Asymlocal}.  
Here  $\phi^\ast$ with $\ast \in \{ \vdash, \dashv, \bot, \top \}$ are given by \eqref{onebasis} and \eqref{otherbasis}. 
We put the constant $\nu(\e,h)$ depending only on $\e$ and $h$ as $\nu(\e,h):= e^{\frac{i\e^2}{2h}\log \e}$. 
\begin{proposition}\label{Asymlocal2} There exists $\mu_0>0$ small enough such that for any $\e$ and $h$ 
with $\frac{\e^2}{h} =: \mu \in (0, \mu_0]$ and one has, uniformly on $h\in (0,h_0]$,  
\begin{align*}
\tilde U [\phi^\vdash](t) &= \widetilde{ \phi^\vdash_0} (t)
\left( 1 + \widetilde{ {\mathcal E}^{\vdash}} \left(t; \e,h\right) \right)&&(t\in I_c(h) \cap \R_+),\\
\tilde U  [\phi^\dashv](t) &= \widetilde{ \phi^\dashv_0} (t)
\left( 1 + \widetilde{ {\mathcal E}^\dashv} (t; \e,h) \right) &&(t\in I_c(h) \cap \R_-),\\
\tilde U [\phi^\bot](t) &= \widetilde{ \pphi^\bot_0} (t)
\left( 1 + \widetilde{ {\mathcal E}^\bot} (t; \e,h) \right) &&(t\in I_c(h) \cap \R_-),\\
\tilde U  [\phi^\top](t) &= \widetilde{ \pphi^\top_0} (t)
\left( 1 + \widetilde{ {\mathcal E}^\top} (t; \e,h) \right) &&(t\in I_c(h) \cap \R_+),
\end{align*}
with
\begin{equation}\label{phi_0} 
\begin{aligned}
\widetilde{ \phi^\vdash_0} (t) &= \frac{\omega(\mu)}{\nu(\e,h)} e^{\frac{i}{h}\int_0^t V(s)ds}\, t^{\frac{i\e^2}{2h}}  
\begin{pmatrix}
-\frac{\e}{2t} \\ 1
\end{pmatrix} ,\\
\widetilde{ \phi^\dashv_0} (t) &= \frac{\omega(\mu)}{\nu(\e,h)} e^{\frac{i}{h}\int_0^t V(s)ds}\,  (-t)^{\frac{i\e^2}{2h}} 
\begin{pmatrix}
-\frac{\e}{2t} \\ 1
\end{pmatrix} ,\\
\widetilde{ \phi^\bot_0} (t) &= {\mathcal C} \left( \frac{\omega(\mu)}{\nu(\e,h)}\right) e^{-\frac{i}{h}\int_0^t V(s)ds}\,  (-t)^{-\frac{i\e^2}{2h}} 
\begin{pmatrix}
1 \\ \frac{\e}{2t}
\end{pmatrix} ,\\
\widetilde{ \phi^\top_0} (t) &= {\mathcal C} \left( \frac{\omega(\mu)}{\nu(\e,h)}\right) e^{-\frac{i}{h}\int_0^t V(s)ds}\,  t^{-\frac{i\e^2}{2h}} 
\begin{pmatrix}
1 \\ \frac{\e}{2t}
\end{pmatrix} ,
\end{aligned}
\end{equation}
where $\omega(\mu) = e^{-\frac{\pi i}{8}} 2^{\frac{1}{4}} \mu^{\frac{i\mu}{2}}$, and 
${\mathcal C}$ stands for the operator of taking its complex conjugate and 
each error $\widetilde{ {\mathcal E}^\ast} (t; \e,h)$ consists of two functions as
$\widetilde{ {\mathcal E}_1^{\ast}} (t; \e,h)
 + \widetilde{ {\mathcal E}_{2}^{\ast}} (t; \e,h)$ satisfying  
$\widetilde{ {\mathcal E}_{1}^{\ast}} (t; \e,h) = \Ord(\frac{\e}{\sqrt{h}})$ uniformly on $t \in I_c(h)$ 
and $\widetilde{ {\mathcal E}_{2}^{\ast}} (t; \e,h) = \Ord(|t|)$ with $\ast \in \{ \vdash, \dashv, \bot, \top\}$. 
\end{proposition}

The proof of Proposition \ref{Asymlocal2} is based on a direct computations 
from Proposition \ref{Asymlocal}. 
In fact, near the crossing point, the solutions of the original equation \eqref{eq1_0} is given by 
the canonical transformation of those of the branching model \eqref{eq1_1}
 (see Proposition \ref{relation_model}, exactly the identity \eqref{phiTOpsi}). 
More precisely, starting from the asymptotic behaviors of the solutions of the local solvable model in a suitable neighborhood (see Proposition \ref{Asymlocal}), 
we obtain \eqref{phi_0} thanks to the change of variables $f$ given by Lemma \ref{local_change}, 
which satisfies $f(t)^2 = 2 \int_0^t V(s) ds$ near the crossing point, 
and by the relationship between three parameters $\e, h$ and $\mu$.



\subsection{{\sl Asymptotic expansions of the exact WKB solutions in some ${\mathcal{O}(\sqrt{h})}$-annulus centered at the crossing point}}\label{wkbsec}

In this subsection, we study the asymptotic expansions of the exact WKB solution near the crossing points.
Assumption \ref{A3} shows that the geometrical setting near each crossing point is the same (see Remark \ref{T_local}).
Then, without loss of generality, we forget the subscript $k$ in all the considered quantities. 
Hereafter, we put $t_k=0$, $\mathcal{S}_k=\mathcal{S}_{\rm local}$, we denote also  the turning point $\zeta_k$ by $\zeta$, 
and the symbol base points $\dl_{k-1,k}$, $\dl_{k,k+1}$ by $r$, $l$ (see  Figure \ref{toyfig}). 
Moreover, we replace  $\zeta, r, l$ (resp. $\bar{\zeta}, \bar{r}, \bar{l}$) with
$\zeta_+, r_+, l_+$ (resp.  $\zeta_-, r_-, l_-$) for the simple notations. 
Then we also express the four WKB solutions \eqref{4wkb_S_k} for simplicity as follows:
\begin{align*}
\psi_\pm^r(t;\e,h) := \psi_\pm(t,\zeta_\pm,r_\pm;h)\quad {\rm and}\quad \psi_\pm^l(t;\e,h) := \psi_\pm(t,\zeta_\pm,l_\pm;h).
\end{align*}

As mentioned in the introduction and explained in Remark \ref{remWKB} in the appendix, 
the approximation of the Wronskian of exact WKB solutions becomes worse close to turning points. 
In particular, when $(\e,h) \to (0,0)$ with ${\scriptstyle \frac{\e^2}{h}} \to 0$, the turning points in ${\mathcal S}_{\rm local}$ 
are very close to the crossing point $t_k=0$.\\

\begin{figure}[h]
\begin{center}
\scalebox{0.35}[0.35]{
\includegraphics{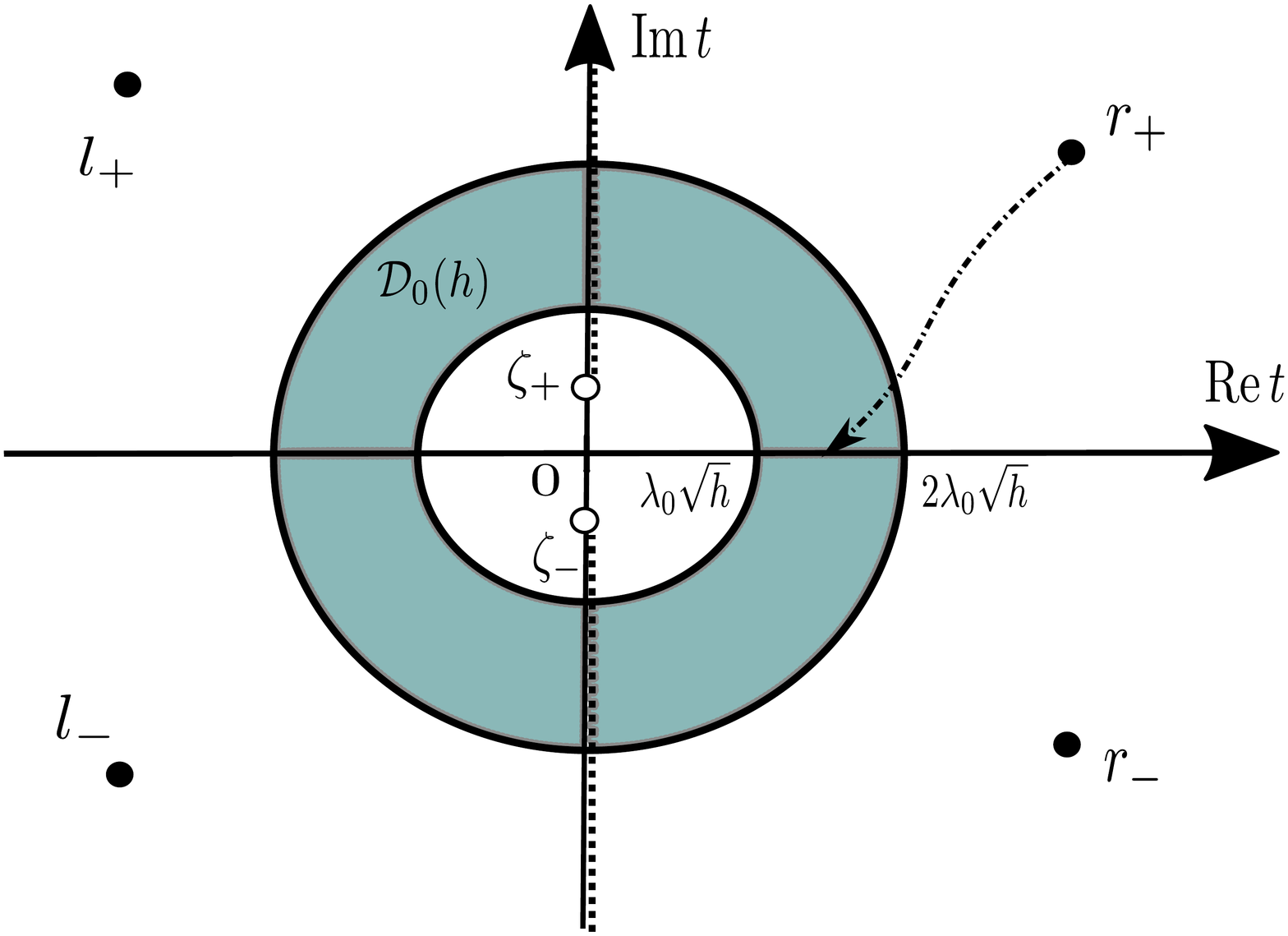}
}
\caption{complex annulus}
\label{annulus}
\end{center}
\end{figure}

Concerning the asymptotic behaviors of the four exact WKB solutions 
$\psi_\pm^r(t;\e,h)$ and $\psi_\pm^l(t;\e,h)$ as $h$ goes to $0$, 
Lemma \ref{wkblem2} in appendix \ref{Rev_WKB} gives 

\begin{align*}
\psi_{\pm}^{r}(t;\e,h) 
 &=\! \frac{1}{2}\!
\begin{pmatrix}
K(z(t))^{-1} \pm K(z(t)) \\
i\{K(z(t))^{-1} \mp K(z(t))\}
\end{pmatrix}\!
\exp\!\left[\pm \frac{z_{\zeta_\pm}(t)}{h} \right]\!
\left(\!1\! +\! {\mathcal O}\left(\!\frac{h}{d(t;\zeta_\pm)}\!\right)\!\right),
\end{align*}
\begin{align*}
\psi_{\pm}^{l}(t;\e,h) 
 &=\! \frac{1}{2}\! 
\begin{pmatrix}
K(z(t))^{-1} \pm K(z(t)) \\
i\{K(z(t))^{-1} \mp K(z(t))\}
\end{pmatrix}\!
\exp\!\left[\pm \frac{z_{\zeta_\pm}(t)}{h} \right]\!
\left(\!1\! +\! {\mathcal O}\left(\!\frac{h}{d(t;\zeta_\pm)}\!\right)\!\right)
\end{align*}
for each $t$ in a suitable subdomain of ${\mathcal S}_{\rm local}$, 
where there exists a canonical curve from each symbol base point toward the origin. 
Notice that, under the regime $\frac{\e^2}{h}\to 0$, the turning points $\zeta_\pm(\e)$ are inside a complex annulus 
${\mathcal D}(h) = \{ t\in \C\ ;\  \lambda \sqrt{h} < |t| < 2\lambda \sqrt{h} \}$ for any positive $\lambda$. 
Then, for $\eta>0$ small enough there exists $\lambda_0>0$ sufficiently large such that 
\begin{equation}\label{error_annu}
\sup_{t \in {\mathcal D}_0(h)} \frac{h}{d(t;\zeta_\pm)} = \Ord (\eta^2)
\end{equation}
uniformly on $\e$ and $h$ with $\frac{\e^2}{h} \to 0$. Here 
\begin{equation}\label{hcompdomain}
 {\mathcal D}_0(h) = \{ t\in \C\ ;\  \lambda_0\sqrt{h} < |t| < 2\lambda_0\sqrt{h} \}.
\end{equation}

From now on, we fix $\lambda_0>0$ large enough such that \eqref{error_annu} holds.
Then, in the regime $\frac{\e^2}{h} \to 0$, we have for $t \in {\mathcal D}_0(h)$ 
\begin{align*}
\psi_{\pm}^{r}(t;\e,h) 
 &= \frac{1}{2}
\begin{pmatrix}
K(z(t))^{-1} \pm K(z(t)) \\
i\{K(z(t))^{-1} \mp K(z(t))\}
\end{pmatrix}
\exp\,\left[\pm \frac{z_{\zeta_\pm}(t)}{h} \right]
\left(1 + {\mathcal O}\left( \eta^2\right)\right),\\
\psi_{\pm}^{l}(t;\e,h) 
 &= \frac{1}{2}
\begin{pmatrix}
K(z(t))^{-1} \pm K(z(t)) \\
i\{K(z(t))^{-1} \mp K(z(t))\}
\end{pmatrix}
\exp\,\left[\pm \frac{z_{\zeta_\pm}(t)}{h} \right]
\left(1 + {\mathcal O}\left(\eta^2\right)\right)
\end{align*}
uniformly with respect to $h \in (0, h_0]$ for some positive $h_0$. 
We denote the leading term of the exact WKB solutions as 
\begin{align*}
\widetilde{ \psi_{\pm}^{r} }(t;\e,h) 
 &:= \frac{1}{2}
\begin{pmatrix}
K(z(t))^{-1} \pm K(z(t)) \\
i\{K(z(t))^{-1} \mp K(z(t))\}
\end{pmatrix}
\exp\,\left[\pm \frac{z_{\zeta_\pm}(t)}{h} \right],\\
\widetilde{ \psi_{\pm}^{l} }(t;\e,h) 
 &:= \frac{1}{2}
\begin{pmatrix}
K(z(t))^{-1} \pm K(z(t)) \\
i\{K(z(t))^{-1} \mp K(z(t))\}
\end{pmatrix}
\exp\,\left[\pm \frac{z_{\zeta_\pm}(t)}{h} \right].
\end{align*}

Throughout this paper we use the following notations:
\begin{equation}\label{hrealdomain}
 I(h):={\mathcal D}_0(h) \cap \R,\quad I^r(h):={\mathcal D}_0(h) \cap \R_+\quad{\rm and}\quad  
 I^l(h):={\mathcal D}_0(h) \cap \R_-.
\end{equation}

\begin{proposition}\label{AsymWKB} 
There exist $\mu_0>0$ and $h_0>0$ small enough such that for any $\e$ and $h$ 
with $\frac{\e^2}{h}  \in (0, \mu_0]$ and $h \in (0,h_0]$, 
each leading term of the exact WKB solutions has the asymptotic behavior 
\begin{align*}
\widetilde{ \psi_+^r }(t;\e,h) &= \psi_{+,0}^r(t;\e,h)
\left( 1 + {\mathcal E}_+^r\left(t;\e,h\right) \right) &&  \text{for} \ \ t\in I^r(h),\\
\widetilde{ \psi_-^r }(t;\e,h) &= \psi_{-,0}^r(t;\e,h)
\left( 1 + {\mathcal E}_-^r\left(t;\e,h\right) \right)  && \text{for} \ \  t\in I^r(h),\\
\widetilde{ \psi_+^l }(t;\e,h) &= \psi_{+,0}^l(t;\e,h)
\left( 1 + {\mathcal E}_+^l\left(t;\e,h\right) \right) && \text{for} \ \ t\in I^l(h),\\
\widetilde{ \psi_-^l }(t;\e,h) &= \psi_{-,0}^l(t;\e,h)
\left( 1 + {\mathcal E}_-^l\left(t;\e,h\right) \right) && \text{for} \ \ t\in I^l(h),
\end{align*}
with
\begin{equation}\label{psi_0}
\begin{aligned}
\psi_{+,0}^r(t;\e,h) &= - \overline{\nu(\e,h)}\, e^{\frac{i}{h}\textstyle\int_0^tV(s)ds}\, t^{\frac{i\e^2}{2h}} 
\begin{pmatrix}
-\frac{\e}{2t} \\ 1
\end{pmatrix},\\
\psi_{-,0}^r(t;\e,h) &= i \nu(\e,h)\, e^{-\frac{i}{h} \textstyle\int_0^tV(s)ds}\, t^{-\frac{i\e^2}{2h}} 
\begin{pmatrix}
1\\ \frac{\e}{2t}
\end{pmatrix},\\
\psi_{+,0}^l(t;\e,h) &= \nu(\e,h)\, e^{-\frac{i}{h}\textstyle\int_0^tV(s)ds}\, (-t)^{-\frac{i\e^2}{2h}} 
\begin{pmatrix}
1\\ \frac{\e}{2t}
\end{pmatrix},\\
\psi_{-,0}^l(t;\e,h) &= i \overline{\nu(\e,h)} e^{\textstyle\frac{i}{h}\int_0^tV(s)ds}\, (-t)^{\frac{i\e^2}{2h}} 
\begin{pmatrix}
-\frac{\e}{2t} \\ 1
\end{pmatrix},
\end{aligned}
\end{equation}
where the error ${\mathcal E}_\pm^{r}(t;\e,h)$ \textnormal{(}resp. ${\mathcal E}_\pm^{l}(t;\e,h)$\textnormal{)} is a function 
satisfying 
$$
{\mathcal E}_\pm^{r}(t;\e,h) =  \Ord\left(\sqrt{h}\right) + \Ord\left(\frac{\e^2}{h}\right)
$$  
as $(\e, h)$ goes to $(0,0)$ with $\frac{\e^2}{h}$ tends to $0$ 
uniformly on $I^r(h)$ \textnormal{(}resp. $I^l(h)$\textnormal{)}.
\end{proposition}

The proof of the above proposition is given in Appendix \ref{proof_WKB}.

\subsection{\sl Correspondence via microsupports}\label{ML connection} 

In this subsection, we deduce some properties of the change of bases between 
the exact WKB solutions and the pull-back solutions of the branching model 
by comparing their microsupports near the crossing point. 
In particular, we obtain the asymptotic behaviors of some special entries of their change of bases, 
which correspond to microlocal solutions with the common microsupports.  

\smallskip
Let us investigate the following relations 
between the exact WKB solutions $\psi_\pm^{r}$, $\psi_\pm^{l}$ and the pull-back solutions of the branching model 
$\tilde U[\phi^\ast]$ with $\ast \in \{ \vdash, \dashv, \bot, \top \}$. 
\begin{equation}\label{relation}
\begin{aligned}
(\psi_+^r\  \psi_-^r) &= ( \tilde U[\phi^\vdash]\  \tilde U[\phi^\dashv] )
\begin{pmatrix}
\rho_{11}^r & \rho_{12}^r\\ \rho_{21}^r & \rho_{22}^r
\end{pmatrix}
= ( \tilde U[\phi^\bot]\  \tilde U[\phi^\top] )
\begin{pmatrix}
\varrho_{11}^r & \varrho_{12}^r\\ \varrho_{21}^r & \varrho_{22}^r
\end{pmatrix},\\[7pt]
(\psi_+^l\  \psi_-^l) &= ( \tilde U[\phi^\vdash]\  \tilde U[\phi^\dashv] )
\begin{pmatrix}
\rho_{11}^l& \rho_{12}^l \\ \rho_{21}^l & \rho_{22}^l
\end{pmatrix}
= ( \tilde U[\phi^\bot]\  \tilde U[\phi^\top] )
\begin{pmatrix}
\varrho_{11}^l & \varrho_{12}^l \\ \varrho_{21}^l & \varrho_{22}^l
\end{pmatrix},
\end{aligned}
\end{equation}
where $\rho_{jk}^{r}, \rho_{jk}^l, \varrho_{jk}^{r}$ and $\varrho_{jk}^l$ are constants depending only on $\e$ and $h$. 
We derive two kinds of the properties on some constants from comparing the microsupports of 
the exact WKB solutions and the pull-back ones of the branching model.

\medskip
We first find that the four constants $\varrho_{21}^r, \rho_{12}^r, \rho_{21}^l$ and $\varrho_{12}^l$ must be zero. 
For $h$ small, the microsupports of WKB solutions of type $\pm$, which is the sign of the phase, satisfy
\begin{align}\label{MS_WKB_0}
{\rm MS}(\psi_\pm) &\subset \Bigl\{ (t,\tau)\in T^*\R \ ;\ \tau = \pm \sqrt{V(t)^2+\e^2} \Bigr\} =: \Lambda_\pm(\e).
\end{align} 
Note that the set $\Lambda_\pm(\e)$ corresponds to 1-dimensional Lagrangian manifold. 
For the definition  of microsupport, MS($\bullet$), and its properties we can consult 
\cite[Appendix A]{Ra96_01}.

\smallskip\noindent
On the other hand, in oder to distinguish the microsupport of the pull-back solutions of the branching model, 
let $\sigma_\star$ $(\star = l,r,u,d)$ be the half-lines in $T^*\R$ given by 
$\sigma_l = \R_- \times \{0\}$, $\sigma_r = \R_+ \times \{0\}$
$\sigma_u = \{0\} \times \R_+$ and $\sigma_d = \{0\} \times \R_-$.
The solutions of the branching model $\phi^\ast$ with $\ast \in \{ \vdash, \dashv, \bot, \top \}$ 
(see \eqref{onebasis} and \eqref{otherbasis}) 
are essentially Heaviside functions. Therefore one sees that, 
\begin{enumerate}
\item[{\rm (i)}] ${\rm MS}(\phi^\vdash)$ is a subset of a neighborhood of $\sigma_r \cup \sigma_u \cup \sigma_d$,
\item[{\rm (ii)}] ${\rm MS}(\phi^\dashv)$ is a subset of a neighborhood of $\sigma_l \cup \sigma_u \cup \sigma_d$,
\item[{\rm (iii)}]  ${\rm MS}(\phi^\bot)$ is a subset of a neighborhood of $\sigma_r \cup \sigma_l \cup \sigma_u$,
\item[{\rm (iv)}]  ${\rm MS}(\phi^\top)$ is a subset of a neighborhood of $\sigma_r \cup \sigma_l \cup \sigma_d$.
\end{enumerate}

The images of the solutions of the branching model by the Fourier integral operator $U_{\frac{\pi}{4}}$ 
can be understood as the microlocal solutions of \eqref{e1} 
from Proposition \ref{relation_model}. So,
the microsupport of  $\tilde U[\phi^\ast]$ with $\ast \in \{ \vdash, \dashv, \bot, \top \}$ 
is the image of ${\rm MS}(\phi^\ast)$ by the canonical transformation $\kappa_{\frac{\pi}{4}}$, 
which is a rotation $\frac{\pi}{4}$ on the phase space, that is 
${\rm MS}(\tilde U[\phi^\ast]) = \kappa_{\frac{\pi}{4}}{\rm MS}(\phi^\ast)$ with $\ast \in \{ \vdash, \dashv, \bot, \top \}$.
One sees that, 
\begin{enumerate}
\item[{\rm (i)}] ${\rm MS}(\tilde U[\phi^\vdash])$ is a subset of a neighborhood of 
$\kappa_{\frac{\pi}{4}}\sigma_r \cup \kappa_{\frac{\pi}{4}}\sigma_u \cup \kappa_{\frac{\pi}{4}}\sigma_d$,
\item[{\rm (ii)}] ${\rm MS}(\tilde U[\phi^\dashv])$ is a subset of a neighborhood of 
$\kappa_{\frac{\pi}{4}}\sigma_l \cup \kappa_{\frac{\pi}{4}}\sigma_u \cup \kappa_{\frac{\pi}{4}}\sigma_d$,
\item[{\rm (iii)}] ${\rm MS}(\tilde U[\phi^\bot])$ is a subset of a neighborhood of 
$\kappa_{\frac{\pi}{4}}\sigma_r \cup \kappa_{\frac{\pi}{4}}\sigma_l \cup \kappa_{\frac{\pi}{4}}\sigma_u$,
\item[{\rm (iv)}] ${\rm MS}(\tilde U[\phi^\top])$ is a subset of a neighborhood of 
$\kappa_{\frac{\pi}{4}}\sigma_r \cup \kappa_{\frac{\pi}{4}}\sigma_l \cup \kappa_{\frac{\pi}{4}}\sigma_d$.
\end{enumerate}
\noindent
Remark that for $\e>0$ small enough, $\Lambda_\pm(\e)$ lie on $\kappa_{\frac{\pi}{4}} \sigma_\star$, $\star\in \{l,r,u,d\}$, 
away from $(0,0)\in T^*\mathbb{R}$.

\smallskip\noindent
Now, let us compare the microsupport of $\psi_+^r$ with those of $\tilde U[\phi^\bot]$ and $\tilde U[\phi^\top]$ on the region with $t>0$. 
While $\psi_+^r$ has the microsupport on $\Lambda_+(\e)$, 
$\tilde U[\phi^\bot]$ has the microsupport on $\kappa_{\frac{\pi}{4}} \sigma_r$, 
which coincides with $\Lambda_+(\e)$ for $\e>0$ small enough,  
however $\tilde U[\phi^\top]$ does the microsupport not only on $\kappa_{\frac{\pi}{4}} \sigma_r$ but also $\kappa_{\frac{\pi}{4}} \sigma_d$. 
This means that the coefficient of $\tilde U[\phi^\top]$, that is $\varrho_{21}^r$, must be zero. 
Similarly we see that the other constants $\rho_{12}^r, \rho_{21}^l$ and $\varrho_{12}^l$ must be zero. 
Moreover this fact implies that there exist proportional relations 
between the exact WKB solutions and the pull-back ones of branching model 
with co-linear coefficients $\varrho_{11}^r$, $\rho_{22}^r$, $\rho_{11}^l$ and $\varrho_{22}^l$. 

\smallskip\noindent 
Second, we also see that the four constants $\rho_{11}^r, \varrho_{22}^r, \varrho_{11}^l$ and $\rho_{22}^l$ 
have asymptotic behaviors which can be deduced from Proposition \ref{AsymWKB} and Proposition \ref{Asymlocal2} 
under a non-adiabatic regime $\frac{\e^2}{h} \to 0$. 
In fact, the leading terms of the WKB solutions \eqref{psi_0} have their microsupports 
included by their Lagrangian manifold $\Lambda_\pm(\e)$.  
On the other hand, the phase factors of the asymptotic behaviors of \eqref{phi_0} and \eqref{psi_0} have the same form: 
$$
\exp \left[ \pm \frac{i}{h} \left( \int_0^t V(s) ds + \frac{\e^2}{2} \log t \right) \right].
$$
This implies that the microsupports of the both microlocal solutions are included by the subset: 
$$
\Bigl\{ (t,\tau)\in T^*\R \ ;\ \tau = \pm \left(V(t) + \frac{\e^2}{2t} \right) \Bigr\}.
$$
\begin{figure}[h]
\begin{center}
\scalebox{0.35}[0.35]{
\includegraphics{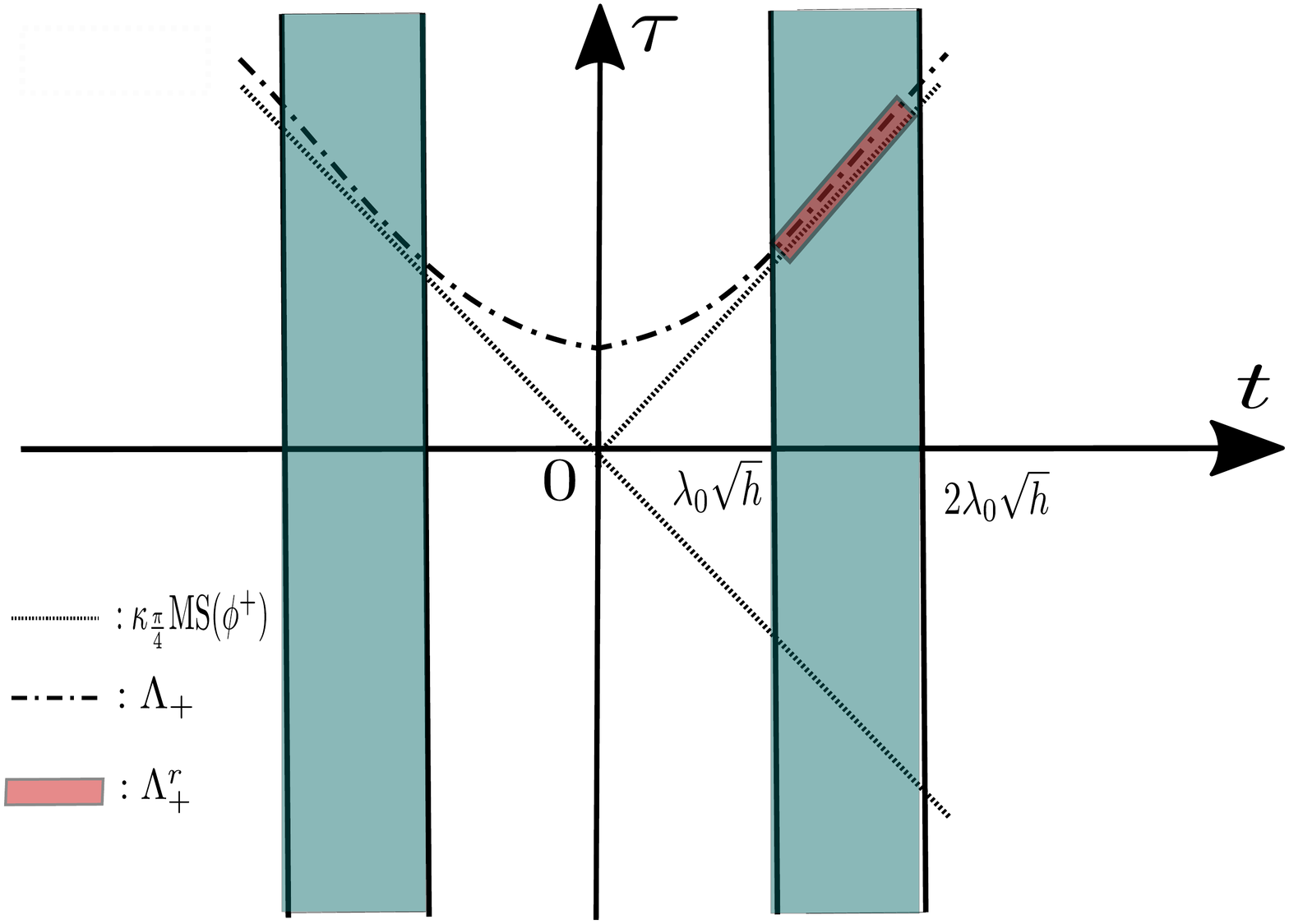}
}
\caption{microsupport on the phase space}
\label{phase}
\end{center}
\end{figure}

We define the subsets of $\Lambda_\pm(\e)$ as
\begin{equation}\label{Lagran_WKB}
\begin{aligned}
\Lambda_\pm^{r}(\e,h) &= \Bigl\{ (t,\tau)\in T^*\R \ ;\ \tau = \pm \sqrt{V(t)^2+\e^2},\quad t \in I^r(h) \Bigr\},\\
\Lambda_\pm^{l}(\e,h) &= \Bigl\{ (t,\tau)\in T^*\R \ ;\ \tau = \pm \sqrt{V(t)^2+\e^2},\quad t \in I^l(h) \Bigr\}.
\end{aligned}
\end{equation}

Notice that, as $\frac{\e^2}{h} \to 0$,  
\begin{equation}
\left| \sqrt{V(t)^2 + \e^2} - \left( V(t) - \frac{\e^2}{2t} \right) \right| \to 0
\end{equation}
uniformly for $t \in I^r(h)$ and $h\in (0, h_0]$.
Hence, taking account of the microsupports of both microlocal solutions, 
and fixing $c$ in \eqref{coer_interval} equal to $\lambda_0$ given in \eqref{error_annu}, 
we see that, for $\frac{\e^2}{h}$ small enough, the followings hold, uniformly for $t \in I^\star(h)$ and $h\in (0, h_0]$,
\begin{equation*}
\begin{aligned}
{\rm MS} (\psi_{+,0}^r)|_{t\in I^r(h)},\  
{\rm MS} (\widetilde{ \phi_0^\vdash} )|_{t\in I^r(h)} &\subset \Lambda_+^{r}(\e,h),\\
{\rm MS} (\psi_{-,0}^r)|_{t\in I^r(h)},\  
{\rm MS}(\widetilde{ \phi_0^\top})|_{t\in I^r(h)} &\subset \Lambda_-^{r}(\e,h),\\
{\rm MS} (\psi_{+,0}^l)|_{t\in I^l(h)},\  
{\rm MS}(\widetilde{ \phi_0^\bot})|_{t\in I^l(h)} &\subset \Lambda_+^{l}(\e,h),\\
{\rm MS} (\psi_{-,0}^l)|_{t\in I^l(h)},\  
{\rm MS}(\widetilde{ \phi_0^\dashv})|_{t\in I^l(h)} &\subset \Lambda_-^{l}(\e,h).
\end{aligned}
\end{equation*}

The above inclusion relations allow us to match on the interval $I^\star(h),\ \star\in\{l,r\}$, 
each leading term in Proposition \ref{Asymlocal2} with 
each corresponding one in Proposition \ref{AsymWKB}. 
Therefore we have  
\begin{lemma}\label{LM_constAsym}
There exist $\mu_0>0$ and $h_0$ small enough such that for any $\frac{\e^2}{h} =: \mu \in (0, \mu_0]$ and $h\in(0, h_0]$ 
the four constants $\rho_{11}^r, \varrho_{22}^r, \rho_{22}^l$ and $\varrho_{11}^l$ in \eqref{relation} 
depending only on $\e$ and $h$ 
have the asymptotic behaviors    
\begin{equation}\label{ConstAsym}
\begin{aligned}
\rho_{11}^r(\e,h) &= \frac{-1}{\omega(\mu)}
 \left( 1 + \Ord(\sqrt{h}) +\Ord\left(\frac{\e}{\sqrt{h}}\right) \right),\\[6pt]
\varrho_{22}^r(\e,h) &= {\mathcal C} \left( \frac{-i}{\omega(\mu)} \right)
 \left( 1 + \Ord(\sqrt{h}) +\Ord\left(\frac{\e}{\sqrt{h}}\right) \right),\\[6pt]
\rho_{22}^l(\e,h) &= \frac{i}{\omega(\mu)}
 \left( 1 + \Ord(\sqrt{h}) +\Ord\left(\frac{\e}{\sqrt{h}}\right) \right),\\[6pt]
\varrho_{11}^l(\e,h) &= {\mathcal C} \left( \frac{1}{\omega(\mu)} \right) 
 \left( 1 + \Ord(\sqrt{h}) +\Ord\left(\frac{\e}{\sqrt{h}}\right) \right),
\end{aligned}
\end{equation}
where $\omega(\mu) = 2^{\frac{1}{4}} e^{-\frac{\pi i}{8}} \mu^{\frac{i \mu}{2}}$ and 
${\mathcal C}$ stands for the operator of taking its complex conjugate.
\end{lemma}

\section{Proof of Theorem \ref{mainthm}}\label{proof_thm}
\setcounter{equation}{0}

The proof of Theorem \ref{mainthm} consists of two parts. 
The first part \ref{Asympt_TM} is to obtain the asymptotic expansion of local transfer matrix $T_{\rm local}(\e,h)$. 
In the second part \ref{SSprodT}, we carry out an algebraic computation of the product of the transfer matrices. 


\subsection{{\sl Asymptotic expansion of the local transfer matrix near the crossing point}}\label{Asympt_TM}


The precise purpose of this subsection is to derive from the preliminaries 
the asymptotic expansion of the transfer matrix $T_{\rm local}(\e,h)$, that is 
\begin{proposition}\label{T_0} There exist $\mu_0>0$ and $h_0>0$ small enough such that for any $\e$ and $h$ 
with $\frac{\e^2}{h} \in (0, \mu_0]$ and $h \in (0,h_0]$ 
the transfer matrix $T_{\rm local}(\e,h)$ has the following asymptotic behavior:
\begin{equation}
T_{\rm local}(\e,h) = 
\begin{pmatrix}
\displaystyle{ e^{i\vartheta}\frac{1}{\bar p} } & \displaystyle{\frac{1}{i} \frac{q}{p}}\\[12pt]
\displaystyle{ \frac{1}{i} \frac{q}{p}} & \displaystyle{ {\mathcal C}\left( e^{i\vartheta}\frac{1}{\bar p} \right) }
\end{pmatrix}
\left(1 + {\mathcal O}\left(\sqrt{h}\right) + {\mathcal O}\left(\frac{\e}{\sqrt{h}}\right)\right),
\end{equation}
where $p, q$ are given by \eqref{pqgamma}, $\vartheta =  \frac{3\pi}{4} + \frac{\e^2}{h}\log \frac{\e^2}{h}$ 
and ${\mathcal C}$ stands for the operator of taking its complex conjugate.
\end{proposition}
Remark that, from \eqref{pq} and $|e^{i\vartheta}|=1$, the determinant of the principal part of $T_{\rm local}$ is $1$ 
and that $\vartheta \to \frac{3\pi}{4}$ when $\frac{\e^2}{h} \to 0$.\\

\smallskip\noindent
Now all of the preparations have been done in the last section, and so let us prove Proposition \ref{T_0}.  
Recalling the relation between the solutions of the branching model $\phi^\ast$ with $\ast \in \{ \vdash, \dashv, \bot, \top \}$  
in Proposition \ref{ChangeBasisNFsol} and the fact that 
$$
\rho_{12}^r = \varrho_{21}^r  = \rho_{21}^l = \varrho_{12}^l = 0,
$$
we have  
\begin{equation}\label{rhoMat}
\begin{aligned}
\begin{pmatrix}
p&-q\\q&-p
\end{pmatrix}
 &=
\begin{pmatrix}
\rho_{11}^r & 0 \\
\rho_{21}^r & \rho_{22}^r 
\end{pmatrix}
\begin{pmatrix}
\varrho_{11}^r & \varrho_{12}^r\\
0 & \varrho_{22}^r
\end{pmatrix}^{-1}
= 
\begin{pmatrix}
\frac{\rho_{11}^r}{\varrho_{11}^r} & -\rho_{11}^r \frac{\varrho_{12}^r}{\varrho_{11}^r\varrho_{22}^r}\\[7pt]
\frac{\rho_{21}^r}{\varrho_{11}^r} & -\rho_{21}^r \frac{\varrho_{12}^r}{\varrho_{11}^r\varrho_{22}^r} + \frac{\rho_{22}^r}{\varrho_{22}^r}
\end{pmatrix},\\
\begin{pmatrix}
p&-q\\q&-p
\end{pmatrix}
&=
\begin{pmatrix}
\rho_{11}^l & \rho_{12}^l \\
0 & \rho_{22}^l 
\end{pmatrix}
\begin{pmatrix}
\varrho_{11}^l & 0\\
\varrho_{21}^l & \varrho_{22}^l
\end{pmatrix}^{-1}
= 
\begin{pmatrix}
\frac{\rho_{11}^l}{\varrho_{11}^l} -\rho_{12}^l \frac{\varrho_{21}^l}{\varrho_{11}^l\varrho_{22}^l} & \frac{\rho_{12}^l}{\varrho_{22}^l}\\[7pt]
-\rho_{22}^l \frac{\varrho_{21}^l}{\varrho_{11}^l\varrho_{22}^l} & \frac{\rho_{22}^l}{\varrho_{22}^l}
\end{pmatrix}.
\end{aligned}
\end{equation}
From Lemma \ref{LM_constAsym}, the asymptotic behaviors of the four constants $\varrho_{11}^r, \rho_{22}^r, \rho_{11}^r$ 
and $\varrho_{22}^l$ are known, hence 
we can solve the unknown constants in \eqref{rhoMat}.   
Consequently, we obtain the connection formulae:
\begin{lemma}\label{co-linear_Coeffi}
The proportional relations between the exact WKB solutions and the pull-back ones of the branching model 
\begin{equation}\label{co-linearity}
\psi^r_+ = \varrho_{11}^r  \tilde U[\phi^\bot], 
\quad \psi^r_- = \rho_{22}^r  \tilde U[\phi^\dashv],
\quad \psi^l_+ = \rho_{11}^l  \tilde U[\phi^\vdash], 
\quad \psi^l_- = \varrho_{22}^l \tilde U[\phi^\top]
\end{equation}
hold with the co-linear coefficients satisfying 
\begin{equation}
\begin{aligned}
\varrho_{11}^r(\e,h) &= \frac{-1}{\omega(\mu)p}  
 \left( 1 + \Ord(\sqrt{h}) +\Ord\left(\frac{\e}{\sqrt{h}}\right) \right),\\[6pt]
\rho_{22}^r(\e,h) &= {\mathcal C} \left(\frac{i}{\omega(\mu)p} \right) 
 \left( 1 + \Ord(\sqrt{h}) +\Ord\left(\frac{\e}{\sqrt{h}}\right) \right),\\[6pt]
\rho_{11}^l(\e,h) &= {\mathcal C} \left(\frac{1}{\omega(\mu)p} \right)
 \left( 1 + \Ord(\sqrt{h}) +\Ord\left(\frac{\e}{\sqrt{h}}\right) \right),\\[6pt]
\varrho_{22}^l(\e,h) &= \frac{-i}{\omega(\mu)p}  
 \left( 1 + \Ord(\sqrt{h}) +\Ord\left(\frac{\e}{\sqrt{h}}\right) \right)
\end{aligned}
\end{equation}
as $(\e, h)$ goes to $(0,0)$ with $\frac{\e^2}{h} =: \mu$ tends to $0$, 
where $\omega(\mu)$ is the same quantity as that given in Proposition \ref{Asymlocal2}.
\end{lemma}

Finally, from Lemma \ref{co-linear_Coeffi}, we can obtain the expression of $T_{\rm local}$ in an algebraic way.
In fact, we can rewrite $(\psi_+^l \psi_-^l) = (\psi_+^r \psi_-^r) T_{\rm local}$ as 
$$
(\tilde U\, {\rm Id})
\bigl(\phi^\vdash \,\phi^\top \bigr)
\begin{pmatrix}\rho_{11}^l & 0\\ 0 & \varrho_{22}^l \end{pmatrix}
= (\tilde U\, {\rm Id})
\bigl(\phi^\bot \,\phi^\dashv \bigr)
\begin{pmatrix}\varrho_{11}^r & 0\\ 0 & \rho_{22}^r \end{pmatrix}
T_{\rm local}.
$$
Remarking that 
$\bigl(\phi^\bot\,\phi^\top\bigr) = \bigl(\phi^\vdash \,\phi^\dashv \bigr)\begin{pmatrix}p&-q\\q&-p\end{pmatrix}$, 
we have
\begin{align*}
T_{\rm local} &= \begin{pmatrix}\varrho_{11}^r & 0\\ 0 & \rho_{22}^r \end{pmatrix}^{-1}\!\!
\begin{pmatrix}p&0\\q&1\end{pmatrix}^{-1}\!\!
\begin{pmatrix}1&-q\\0&-p\end{pmatrix}
\begin{pmatrix}\rho_{11}^l & 0\\ 0 & \varrho_{22}^l \end{pmatrix}
 = \left(
\begin{matrix}
\displaystyle{\frac{\rho_{11}^l}{\varrho_{11}^r} \frac{1}{p}}
 & -\displaystyle{\frac{\varrho_{22}^l}{\varrho_{11}^r} \frac{q}{p}}\\[12pt]
 -\displaystyle{\frac{\rho_{11}^l}{\rho_{22}^r} \frac{q}{p}}
 & \displaystyle{-\frac{\varrho_{22}^l}{\rho_{22}^r} \frac{1}{\bar{p}}}
\end{matrix}
\right).
\end{align*}
Here we used \eqref{pq}. Moreover, applying Lemma \ref{co-linear_Coeffi} for these constants, we have 
\begin{align*}
T_{\rm local} &= - \left(
\begin{matrix}
\displaystyle{\frac{\omega}{\bar{\omega}} \frac{1}{\bar{p}}}
 & \displaystyle{i\frac{q}{p}}\\[12pt]
\displaystyle{i \frac{q}{p}}
 & \displaystyle{\frac{\bar{\omega}}{\omega}  \frac{1}{p}}
\end{matrix}
\right)
\left( 1 + \Ord(\sqrt{h}) +\Ord\left(\frac{\e}{\sqrt{h}}\right) \right)
\end{align*}
as $(\e, h)$ goes to $(0,0)$ with $\frac{\e^2}{h}$ tends to $0$. Recalling that 
$\omega (\bar{\omega})^{-1} = e^{i \vartheta}$ with $\vartheta = -\frac{\pi}{4} + \frac{\e^2}{h} \log \frac{\e^2}{h}$,
the proof of Proposition \ref{T_0} is complete.


\subsection{{\sl Product of the transfer matrices}}\label{SSprodT}

In this subsection, we first derive the asymptotic behavior of each transfer matrix $T_k(\e,h)$ from the result 
of the previous one (see Proposition \ref{T_0}) and after compute the product of these transfer matrices. 

Taking account the translation and the scaling: $t \mapsto \sqrt{v_k}(t-t_k)$, 
we obtain $T_k(\e,h)$ as follows: 
\begin{equation}\label{Tk}
T_k(\e,h) = 
\begin{pmatrix}
\displaystyle{ e^{i\vartheta} \frac{1}{\overline{p_k} } } & \displaystyle{\frac{1}{i}\frac{q_k}{p_k}}\\[12pt]
\displaystyle{\frac{1}{i}\frac{q_k}{p_k}} & \displaystyle{ {\mathcal C} \left(e^{i\vartheta} \frac{1}{\overline{p_k} }\right) }
\end{pmatrix}
\left(1 + \Ord(\sqrt{h}) + {\mathcal O}\left(\frac{\e}{\sqrt{h}}\right)\right)
\end{equation}
as $(\e,h)\to (0,0)$ with $\frac{\e^2}{h} \to 0$, where $\vartheta$ is given in Proposition \ref{T_0} and 
\begin{equation}\label{p_k_q_k}
p_k= \gamma_k e^{\frac{\pi}{4v_k}\frac{\e^2}{h}}, \
q_k= \gamma_k e^{-\frac{\pi}{4v_k}\frac{\e^2}{h}},\quad
\gamma_k = \frac{1}{i} \sqrt{\frac{v_k h}{\pi \e^2}}\! \left(\frac{\e^2}{h}\right)^{-\frac{i}{2v_k} \frac{\e^2}{h}}\!\!\!
\Gamma\left( 1-\frac{i}{2v_k} \frac{\e^2}{h} \right).
\end{equation}
Now we have gotten the asymptotic behaviors of all kinds of transfer matrices $T_k$, $T_{k,k+1}$, $T_r$ and $T_l$
(see \eqref{Tk}, \eqref{T_kk} and \eqref{r-inf}). 
From Proposition \ref{S-represent}, we rewrite the scattering matrix $S(\e,h)$ by means of the notation
${\mathcal T}_k = T_k T_{k,k+1}$ for $k =0,1,2,\ldots, n$ with $T_0 = {\rm Id}$ as 
\begin{align}\label{Smatrix_prodT}
S(\e,h) = T_{0,1}(\e,h) T_1(\e,h) T_{1,2}(\e,h)  \cdots T_n(\e,h) T_{n,n+1}(\e,h)
 = \prod_{k=0}^n {\mathcal T}_k(\e,h), 
\end{align}
where $T_{0,1}(\e,h) = T_r(\e,h)^{-1}$ and $T_{n,n+1}(\e,h) = T_l(\e,h)$. 
In order to understand the structure of $S(\e,h)$, we make use of elementary notations $D_1$, $D_2$, $N_1$, $N_2$ 
which are introduced in Appendix \ref{Alg_lemmas} (see \eqref{M2ring}). 
We see that 
\begin{align*}
T_{k.k+1} = a_k D_1 + \overline{a_k} D_2,\qquad 
T_k = b_k D_1 + \overline{b_k} D_2 + c_k N_1 + c_k N_2,
\end{align*}
where $a_k$ is the same as our notations of the actions (see \eqref{a_k}, \eqref{a_r_l}), 
and $b_k = e^{i\vartheta}\frac{1}{\overline{p_k}}$, $c_k =\frac{1}{i}\frac{q_k}{p_k}$ 
given in \eqref{Tk}, \eqref{p_k_q_k}. 
Moreover we regard $T_{0,1}$ and $T_{n,n+1}$ as 
\begin{align*}
T_{0,1} = T_r^{-1} = \begin{pmatrix} -1 & 0\\ 0 & i\end{pmatrix} \begin{pmatrix} a_0 & 0\\ 0 & \overline{a_0}\end{pmatrix},\quad 
T_{n,n+1} = T_l = \begin{pmatrix} a_n & 0\\ 0 & \overline{a_n}\end{pmatrix} \begin{pmatrix} -1 & 0\\ 0 & -i\end{pmatrix} 
\end{align*}
with $a_0 = a_r^{-1}$ and $a_n = a_l$. 
From the definitions of $p_k, q_k$ and the fact $|e^{i\vartheta}| =1$, we have
\begin{align*}
|b_k|^2 = \left|\frac{1}{p_k}\right|^2 &= \frac{\pi}{v_k} \frac{\e^2}{h}
\left(1+{\mathcal O}\left(\frac{\e^2}{h}\right)\right),\\
|c_k|^2 = \left|\frac{q_k}{p_k}\right|^2 &= 1 - \frac{\pi}{v_k} \frac{\e^2}{h}
\left(1+{\mathcal O}\left(\frac{\e^2}{h}\right)\right),
\end{align*}
as $(\e,h)\to (0,0)$ with $\frac{\e^2}{h} \to 0$. 

\medskip\noindent
Notice that our setting satisfies the hypothesis of Lemma \ref{alg_lem1}. 
We can understand, from Lemma \ref{alg_lem1}, the asymptotic behavior of the scattering matrix, 
in particular whether diagonals or off-diagonals are dominants according to the parity of the number $n$. 

\medskip\noindent
In order to obtain the prefactor $C_n(h)$ of the transition probability $P(\e,h)$ 
in Theorem \ref{mainthm}, we must take account of not only the dominant term $\Ord(1)$ of the transfer matrix $T_k$ 
but also the subdominant $\Ord(|b_k|) = \Ord(\frac{\e^2}{h})$. 
In fact, when $n$ is odd, it is complicated to compute $|s_{21}(\e,h)|^2$ itself directly. 
However, thanks to the unitary property $|s_{11}|^2 + |s_{21}|^2 = 1$, 
it is enough to compute $|s_{11}|^2$ instead of $|s_{21}|^2$. 
Consequently, it can be reduced to compute $|\tau_n(\bm{a},\bm{b},\bm{c})|^2$ 
defined by \eqref{def_tau} whose expression is given by Lemma \ref{tau^2compu3}. Then, 
\begin{equation*}
\begin{aligned}
|\tau_n|^2 &= \left( \prod_{l=0}^n |a_l|^2 \right) \left( \prod_{l=1}^n |c_l|^2 \right) \sum_{k=1}^n 
\left| \frac{b_k}{c_k} \right|^2 \\
 &\quad + 2 \left( \prod_{l=1}^n |c_l|^2 \right) {\rm Re}\, \sum_{k=2}^{n} 
 \Biggl[
 \left( \prod_{l=k}^n |a_l|^2 \right) \left( {\mathcal C}^{(k)} b_k\right) \left( {\mathcal C} \frac{1}{c_k} \right)\\
 &\hspace{4cm} \sum_{j=1}^{k-1}
 \left( \prod_{l=0}^{j-1} |a_l|^2 \right) \left( \prod_{l=j-1}^{k-2} \left({\mathcal C}^{(l)} a_{l+1}^2 \right)\right)  
\left( {\mathcal C}^{(j-1)} b_{j}\right) \frac{1}{c_{j}}
 \Biggr].
\end{aligned}
\end{equation*}

\smallskip\noindent
For the product of the actions between crossing points, we know 
\begin{equation}\label{Act_cancel}
\begin{aligned}
a_{k-1}a_k &= \exp \left[ \frac{i}{2h} \Bigl(A_{k-1}(\e) - A_{k+1}(\e) + R_{k-1}(\e) + R_{k}(\e) \Bigr) \right],\\
a_{k-1}\overline{a_k} &= \exp \left[ \frac{i}{2h} \Bigl(A_{k-1}(\e) + \overline{A_{k+1}}(\e) - 2{\rm Re}\,A_k(\e)
 + R_{k-1}(\e) - R_{k}(\e) \Bigr) \right].
\end{aligned}
\end{equation}

\medskip\noindent
Taking account of the actions coming from Jost solutions at $\pm \infty$, 
we note that all of actions cancel or become into their real parts. 
For any natural number $0 \leq k\leq n$, one sees that, from the above fact, 
\begin{align*}
\prod_{l=k}^n |a_l|^2 &= 1+{\mathcal O}\left(\frac{\e^2}{h}\right),
\end{align*}
and also has for any $j,k \in \{0, 1,\ldots, n\}$,
$$
c_{j} \overline{c_k} = 1+{\mathcal O}\left(\frac{\e^2}{h}\right)
$$
as $(\e,h)\to (0,0)$ with ${\scriptstyle \frac{\e^2}{h} }\to 0$. 

\medskip\noindent
Hence \eqref{tau^2compu1} can be reduced to 
\begin{align}\label{tau^2compu5}
|\tau_n|^2 &= \sum_{k=1}^n \frac{\pi \e^2}{v_k h}  
+ 2 {\rm Re}\, \sum_{k=2}^n \left( {\mathcal C}^{(k)} b_k\right)
 \sum_{j=1}^{k-1}  \left( \prod_{l=j-1}^{k-2} \left({\mathcal C}^{(l)} a_{l+1}^2 \right)\right)  
\left( {\mathcal C}^{(j-1)} b_{j}\right),  
\end{align}
with modulo $\Ord(\frac{\e^4}{h^2})$. Let us carry on the computation of the second summation. 
\begin{align}\nonumber
 & \sum_{k=2}^n \left( {\mathcal C}^{(k)} b_k\right)
 \sum_{j=1}^{k-1}  \left( \prod_{l=j-1}^{k-2} \left({\mathcal C}^{(l)} a_{l+1}^2 \right)\right)  
\left( {\mathcal C}^{(j-1)} b_{j}\right)\\ \label{tau^2compu6}
 =\ & \sum_{k=2}^n 
 \sum_{j=1}^{k-1}  \left\{ {\mathcal C}^{(j+1)} \prod_{l=0}^{k-1-j} \left({\mathcal C}^{(l)} a_{j+l}^2 \right) \right\}  
\left\{ {\mathcal C}^{(j)} \left( \overline{b_{j+1}} \left( {\mathcal C}^{(k-j)} b_k\right) \right)\right\}.
\end{align}

\medskip\noindent\noindent
Remark that from definition of $b_k$ the followings hold.
\begin{equation}\label{pp_formula}
\begin{aligned}
b_jb_k &= 
 \frac{\pi i}{\sqrt{v_jv_k}}\frac{\e^2}{h} 
\exp \left[
-\frac{i}{2}\left(\frac{1}{v_j}+\frac{1}{v_k} - 4\right) \left(\frac{\e^2}{h}\log \frac{\e^2}{h}\right)
\right]
\left(1+{\mathcal O}\left(\frac{\e^2}{h}\right)\right),\\
b_j \overline{b_k} &= 
\frac{\pi}{\sqrt{v_jv_k}}\frac{\e^2}{h} 
\exp \left[
-\frac{i}{2}\left(\frac{1}{v_j}-\frac{1}{v_k}\right) \left(\frac{\e^2}{h}\log \frac{\e^2}{h}\right)
\right]
\left(1+{\mathcal O}\left(\frac{\e^2}{h}\right)\right),
\end{aligned}
\end{equation}
as $(\e,h)\to (0,0)$ with $\frac{\e^2}{h} \to 0$.

\medskip\noindent
By using the above formulae \eqref{Act_cancel} and \eqref{pp_formula}, we know
\begin{align*}
 &{\mathcal C}^{(j+1)} \prod_{l=0}^{k-1-j} \left({\mathcal C}^{(l)} a_{j+l}^2 \right) 
=  {\mathcal C}^{(j+1)}\left(  a_j^2 \overline{a_{j+1}^2} {a_{j+2}^2} \overline{a_{j+3}^2} \cdots \left( {\mathcal C}^{(k-1-j)} a_{k-1}^2 \right) \right)\\
 =\ & {\mathcal C}^{(j+1)}\Biggl( \exp \left[ -\frac{1}{h} {\rm Im}\, {\widetilde A}_{j,k} \right] \\
 & \hspace{12mm} \times \exp \left[ \frac{i}{h}\left(\sum_{l=0}^{k-1-j} (-1)^l R_{j+l} + 2\sum_{l=1}^{k-1-j} (-1)^l {\rm Re}\, A_{j+l}
 + {\rm Re}\, {\widetilde A}_{j,k} \right)\right] \Biggr),\\
 =\ & \exp \left[ (-1)^{j+1} \frac{i}{h}\left(\sum_{l=0}^{k-1-j} (-1)^l {\mathcal R_{j+l}} \right)\right] 
\left( 1 + \Ord\left( \frac{\e^2}{h} \right) \right),
\end{align*}
where ${\widetilde A}_{j,k}:=A_{j}+(-1)^{k-j}{\mathcal C}^{(k-1-j)}A_k$ and ${\mathcal R_{j}} := 2\int_{t_{j+1}}^{t_j} |V(t)|\,dt$ satisfying $R_j(\e) = {\mathcal R}_j + \Ord(\e^2)$, 
and also
\begin{align*}
  {\mathcal C}^{(j)} \left( \overline{b_{j+1}} \left( {\mathcal C}^{(k-j)} b_k\right) \right)
=  \frac{\pi}{\sqrt{v_j v_k} } \frac{\e^2}{h} 
\exp \left[ (-1)^{j+1} \kappa_{k-j} \frac{\pi}{2}i \right]\!\!
\left(\! 1\! +\! \Ord\left(\! \frac{\e^2}{h}\log \frac{\e^2}{h} \!\right) \!\right),
\end{align*}
where $\kappa_{l} = \frac{1-(-1)^{l}}{2}$ as $(\e,h)\to (0,0)$ with $\frac{\e^2}{h} \to 0$.

\medskip\noindent
By means of these computations, we can derive the prefactor $C_n(\e,h)$ in the main theorem 
from \eqref{tau^2compu6} as follows. 
\begin{align*}
 & {\rm Re}\, \sum_{k=2}^n \sum_{j=1}^{k-1}  \left\{ {\mathcal C}^{(j+1)} \prod_{l=0}^{k-1-j} \left({\mathcal C}^{(l)} a_{j+l}^2 \right) \right\}  
\left\{ {\mathcal C}^{(j)} \left( \overline{b_{j+1}} \left( {\mathcal C}^{(k-j)} b_k\right) \right)\right\},\\
 =\ & {\rm Re}\, \sum_{k=2}^n \sum_{j=1}^{k-1} \frac{\pi}{\sqrt{v_j v_k} } \frac{\e^2}{h} 
\exp \left[ (-1)^{j+1} i \left( \frac{1}{h} \sum_{l=0}^{k-1-j} (-1)^l {\mathcal R_{j+l}} +  \kappa_{k-j} \frac{\pi}{2}\right) \right]\\
 & \hspace{82mm}\times \left( 1 + \Ord\left( \frac{\e^2}{h}\log \frac{\e^2}{h} \right) \right),\\
 \end{align*}

 \begin{align*}
 =\ &  \sum_{k=2}^n \sum_{j=1}^{k-1} \frac{\pi}{\sqrt{v_j v_k} } \frac{\e^2}{h} 
\cos \left[  \frac{1}{h} \sum_{l=j}^{k-1} (-1)^{l-j} {\mathcal R_{l}} +  \kappa_{k-j} \frac{\pi}{2} \right]
\left( 1 + \Ord\left( \frac{\e^2}{h}\log \frac{\e^2}{h} \right) \right),\\
=\ &  \sum_{k=2}^n \sum_{j=1}^{k-1} \frac{\pi}{\sqrt{v_j v_k} } \frac{\e^2}{h} 
\cos \left[  \frac{1}{h} \sum_{l=j}^{k-1} (-1)^{l} {\mathcal R_{l}} +  \frac{(-1)^j-(-1)^k}{2} \frac{\pi}{2} \right]\\
 & \hspace{82mm}\times \left( 1 + \Ord\left( \frac{\e^2}{h}\log \frac{\e^2}{h} \right) \right),
\end{align*}
as $(\e,h)\to (0,0)$ with $\frac{\e^2}{h} \to 0$. \\

\noindent
Notice that the error $\Ord( (\frac{\e^2}{h})^2 \log \frac{\e^2}{h})$ is smaller than $\Ord( (\frac{\e^2}{h})^{\frac{3}{2}})$ 
and $\overset{k-1}{\underset{l=j}{\sum}} (-1)^l {\mathcal R_{l}} = \displaystyle 2\int_{t_{k}}^{t_{j}}V(t)dt$. 
Hence we obtain Theorem \ref{mainthm}.

\bigskip

\section{Proof of Theorem \ref{thmWKB}}\label{proofWKBthm4}

In the adiabatic case (${\scriptstyle \frac{h}{\e^2}} \to 0$), exact WKB solutions are valid  
even near crossing points, while in the non-adiabatic case (${\scriptstyle \frac{\e^2}{h}} \to 0$), those are not.   
Therefore the exact WKB method reviewed in Appendix \ref{Rev_WKB} 
provides us the asymptotic behavior of the local transfer matrix near each crossing point.  
Applying the Wronskian formula (see Lemma \ref{wkblem3}) to the expression of $T_k(\e,h)$ (see \eqref{T_kWron}), 
we can deduce the following asymptotic formula. 
\begin{lemma}\label{adiabatic_T_k}
For $k =1,2,\ldots, n$, we have
\begin{equation}\label{T_k_asympt_adiabatic}
T_k(\e,h) = 
\begin{pmatrix}
1 +  {\mathcal O}\left( \frac{h}{\e^2}\right) & (-1)^{k-1} i e^{\frac{i}{h}A_k(\e)}(1+{\mathcal O}(h))\\[7pt]
(-1)^{k-1} i e^{\frac{i}{h}A_k(\e)}(1+{\mathcal O}(h)) & 1 +  {\mathcal O}\left( \frac{h}{\e^2}\right) 
\end{pmatrix}
\end{equation}
as $(\e,h) \to (0,0)$ and ${\scriptstyle \frac{h}{\e^2}} \to 0$, where $A_k(\e)$ is given by \eqref{Actionk}. 
\end{lemma}
The proof of this lemma was done explicitly under a generic one-crossing model in \cite[Section 4]{Wa06_01}. 
Now, we can prove Theorem \ref{thmWKB} in similar way to the subsection \ref{SSprodT} 
by means of the same notations. 
Recalling \eqref{Smatrix_prodT}, 
the scattering matrix is expressed by the product of the matrix ${\mathcal T}_k := T_k T_{k,k+1}$, 
where $T_{k,k+1}$ is as in \eqref{T_kk}. 
We can write ${\mathcal T}_k$ as 
\begin{equation*}
{\mathcal T}_k = a_kb_k D_1 + \overline{a_k} b_k D_2 + a_kc_k N_1 + \overline{a_k}c_k N_2, 
\end{equation*} 
where the matrices $D_1, D_2, N_1$ and $N_2$ are given by \eqref{M2ring} and 
$a_k$ is the same as in Appendix \ref{S_matrix_section}. 
Here, in the adiabatic regime, $b_k$ and $c_k$ have asymptotic behaviors 
\begin{align}\label{asymp_prop_bc}
b_k = 1 +  {\mathcal O}\left( \frac{h}{\e^2}\right),\quad c_k = (-1)^{k-1} i e^{\frac{i}{h}A_k(\e)}(1+{\mathcal O}(h))
\end{align}
as $(\e,h) \to (0,0)$ and ${\scriptstyle \frac{h}{\e^2}} \to 0$. 
The crucial point is that the modulus of off-diagonal term $c_k$ is exponentially decaying. 
In fact, recalling that 
${\rm Im}\, A_k(\varepsilon)$ is positive and of $\Ord(\e^2)$ as $\e \to 0$, 
we see that 
$|c_k| = \Ord(e^{-\frac{\alpha(\e)}{h}})$, 
where $\alpha (\varepsilon) := \underset{k\in\mathcal{K}}{\min}\, 
\big(\textnormal{Im}\,A_k(\varepsilon)\big)$ with $\mathcal{K}$ the set of $k\in\{1,2,\cdots,n\}$ 
which attains $\max\{v_1,\cdots,v_n\}$. 
Here $v_k=|V'(t_k)|>0$ for $k=1,\ldots,n$. \\

\noindent Taking account of the above properties, we can compute the asymptotic behavior of the product of ${\mathcal T}_k$.  
For $m = 1, \dots, n$, we introduce the operation ${\mathcal G}_m$ as 
$$
{\mathcal G}_m (\prod_{j =1}^{n} a_{j} b_{j}) = 
(\prod_{j=1}^{m-1} \overline{a_{j}}b_{j}) (a_{m}c_{m}) (\prod_{j=m+1}^{n} a_{j}b_{j}),
$$
where, by convention, the first (resp. third) factor of RHS is 1 when $m=1$ (resp. $m = n$). 
Then, by induction, we get
\begin{align*}
\prod_{k=1}^n {\mathcal T}_k = 
\begin{pmatrix}
{\displaystyle \prod_{k=1}^n a_{k} b_{k} } & 
{\displaystyle \sum_{m=1}^{n} {\mathcal G}_m (\prod_{k=1}^n \overline{a_{k}} b_{k}) }\\
{\displaystyle \sum_{m=1}^{n} {\mathcal G}_m (\prod_{k=1}^n a_{k} b_{k}) } & 
{\displaystyle \prod_{k=1}^n \overline{a_{k}} b_{k} }
\end{pmatrix} 
+ \Ord(e^{-\frac{2\alpha(\e)}{h}}). 
\end{align*}
Then the transition probability $|s_{21}|^2$ is equal to 
\begin{equation}\label{computation_s21}
\begin{aligned}
 &\left| \overline{a_0} (1+ \Ord(h))\sum_{m=1}^{n} {\mathcal G}_m (\prod_{k=1}^n a_{k} b_{k}) 
+ \Ord(e^{-\frac{2\alpha(\e)}{h}}) \right|^2\\ 
 &\qquad = \left| \overline{a_0} \sum_{m=1}^{n} {\mathcal G}_m (\prod_{k=1}^n a_{k} b_{k})  \right|^2 
+ \Ord(e^{-\frac{3\alpha(\e)}{h}}),
\end{aligned}
\end{equation}
since for each $m$ the term ${\mathcal G}_m (\overset{n}{\underset{k=1}{\prod}} a_{k} b_{k})$ 
contains $c_m$ whose order is $\Ord(e^{-\frac{\alpha(\e)}{h}})$. 
Here $a_0 = e^{-\frac{i}{2h}\left(A_1(\e)-A_{r}(\e)+2\lambda_rt_1\right)}$, (see \eqref{r-inf} and also \eqref{a_r_l}). 
Now we can deduce the main term of the transition probability from the actions.
According to the formulas of the products of actions \eqref{Act_cancel} and 
$a_n = e^{\frac{i}{2h}\left(A_n(\e)-A_{l}(\e)+2\lambda_lt_n\right)}$, we have, for $m = 1, \ldots, n$, 
\begin{align}\nonumber
 & \overline{a_0a_1\cdots a_{m-1}} a_{m} a_{m+1}\cdots a_{n}\\ \nonumber
= \ & e^{ \frac{i}{2h}\bigl[ (\overline{A_1} - A_r  + 2\lambda_r t_1)
 - \bigl(\overset{m-1}{\underset{k=1}{\sum}}\overline{A_k} - \overline{A_{k+1}} + R_k\bigr)
 + \bigl(\overset{n-1}{\underset{k=m}{\sum}} A_k - A_{k+1} + R_k\bigr)
 + (A_n - A_l + 2\lambda_l t_n) \bigr]  }\\ \nonumber
= \ & e^{ \frac{i}{2h}\bigl[ - A_r  + 2\lambda_r t_1  - A_l + 2\lambda_l t_n \bigr]  }
 e^{ \frac{i}{2h}\bigl[ 2 {\rm Re}\, A_{m}  -\overset{m-1}{\underset{k=1}{\sum}} R_k
 + \overset{n-1}{\underset{k=m}{\sum}} R_k \bigr] }\\ \label{calucu_a}
= \ & e^{ \frac{i}{2h}\bigl( 2\lambda_r t_1 + 2\lambda_l t_n  - A_r  - A_l + \overset{n-1}{\underset{j=1}{\sum}} R_j\bigr)  }
 e^{ \frac{i}{h}\bigl( {\rm Re}\, A_{m}  -\overset{m-1}{\underset{j=1}{\sum}} R_j \bigr)  }.
\end{align}
The prefactor of \eqref{calucu_a} is independent of $m$ and its modulus is 1. 
By using the asymptotic behaviors \eqref{asymp_prop_bc}, we get, for $m = 1, \ldots, n$,
\begin{equation}\label{calucu_bc}
b_1 \cdots b_{m-1} c_{m} b_{m+1}\cdots b_{n} 
=  (-1)^{m-1} i e^{\frac{i}{h} A_{m}(\e)}\left(1+{\mathcal O}\left( \frac{h}{\e^2} \right)\right).
\end{equation}
Combining \eqref{calucu_a} and \eqref{calucu_bc}, we can compute the summation \eqref{computation_s21} as
\begin{align*}
 & \left| \sum_{m=1}^{n} (-1)^m e^{ \frac{i}{h}\bigl( A_{m}(\e) + {\rm Re}\, A_{m}
 - \overset{m-1}{\underset{j=1}{\sum}} R_j \bigr)  } \right|^2
\left(1 + {\mathcal O}\left( \frac{h}{\e^2} \right)\right) +  \Ord\left(e^{-\frac{3\alpha(\e)}{h}}\right).
\end{align*}
Considering once more the magnitude of the imaginary part of the action ${\rm Im}\, A_m(\e)$ 
and comparing them, 
we obtain the asymptotic formula \eqref{thmWKBformula} in Theorem \ref{thmWKB}.

\begin{appendix}
 
\section{{\sl Exact WKB approach}}\label{Rev_WKB}

This appendix is devoted to a quick review of the exact WKB method.
In particular, we fix the notations used in the paper.

\subsection{{ Construction of the exact WKB solutions}}\label{constWKB}

In this subsection, we recall the construction of exact WKB solutions specific to our situation with the parameter $\varepsilon$ fixed
for the moment. 
This construction was initiated by G\'erard-Grigis (see \cite{GeGr88_01}) 
and developed to a first order $2\times 2$ system by Fujii\'e-Lasser-N\'ed\'elec (see \cite{FuLaNe09_01}).\\  

Set $\psi(t;h):= \frac{1}{2}\begin{pmatrix}1&i\\i&1\end{pmatrix}\phi(t;h)$,
where $\phi(t;h)$ is the solution of \eqref{e1}.
Then the original equation \eqref{e1} can be reduced to 
 the following first order $2\times 2$ system:
\begin{equation}\label{wkbModel}
\frac{h}{i} \frac{d}{dt} \phi(t;h) = 
\begin{pmatrix}
0 & \alpha(t)\\ -\beta(t) & 0
\end{pmatrix}
\phi(t;h),
\end{equation}
where $\alpha(t) = -iV(t) - \e$ and $\beta(t) = -iV(t) + \e$.
One sees that the equation \eqref{wkbModel} is a natural extension of the Schr\"{o}dinger equation 
by taking $\alpha(t)=1$ and $\beta(t) = V(t)-E$.

\smallskip\noindent
We treat this equation on a simply connected domain ${\mathcal S}\subset \C$ given in  \ref{A1} 
and define for any fixed point $a\in {\mathcal S}$
$$
z_a(t) = \int_a^t \sqrt{\alpha(s)\beta(s)}\,ds = i \int_a^t \sqrt{V(s)^2 + \e^2}\, ds,
$$
where the branch of the integrand $\sqrt{V(t)^2 + \e^2}$ is taken $\e$ at vanishing points of $V(t)$
i.e., $t= t_k$ for $k = 1,\ldots,n$. 
One sees that $z_a(t)$ satisfies so-called eikonal equation of \eqref{wkbModel}. 
Notice that for any $a, \tilde a \in {\mathcal S}$ one has 
\begin{equation}\label{z_affine}
z_a(t) = z_{\tilde a}(t) + \int_a^{\tilde a} \sqrt{\alpha(s)\beta(s)} ds.
\end{equation}

We denote by $\Lambda$ a set of turning points which are zeros of $\alpha(t)\beta(t)$, and 
by $\widetilde {\mathcal S}$ the simply connected domain ${\mathcal S}\setminus \Lambda$. 
Remark that the mapping $z_a$ is bijective from $\widetilde {\mathcal S}$ to $z_a(\widetilde {\mathcal S})$. 
From \ref{A1}, we can find a suitable small constant $c$ and define the strip domain 
${\mathcal S}_{\rm bdd} =\{ t\in {\mathcal S}\,; \ |\text{Im}\, t| < c \}$ such that
${\mathcal S}_{\rm bdd} \cap \Lambda  
= \{ \zeta_k, \overline{\zeta_k}\, ;\, k= 1,\ldots, n \}$ by the assumption \ref{A3} 
and the Rouch\'e theorem.
We make a branch cut from each $\zeta_k$ (resp. $\overline{\zeta_k}$) 
in the direction parallel to the imaginary axis with the positive (resp. negative) 
imaginary part (see Figure \ref{S_box}).
Note that under this choice of the branch cut the
whole real axis is included in the corresponding simply connected subdomain of 
$\widetilde {\mathcal S}_{\rm bdd}:=\widetilde {\mathcal S}\cap {\mathcal S}_{\rm bdd}$.
This fact permits us to know that $\pm \text{Re}\,z_a(t)$ increase as $\pm \text{Im}\,t$ decrease.\\

\noindent
In this context, we can consider the solution of \eqref{wkbModel}, 
$\phi(t;h)$, as a function of the variable $z$ by setting
$\phi_\pm(t;h) = e^{\pm \frac{z}{h}} M_\pm(z) w_\pm(z;h)$, with  
$$
M_\pm(z) = \begin{pmatrix}
K(z)^{-1} & K(z)^{-1}\\
\mp i K(z) & \pm i K(z)
\end{pmatrix}, \quad 
K(z(t)) = \left(\frac{\beta(t)}{\alpha(t)}\right)^{\frac{1}{4}} 
= \left(\frac{-iV(t) + \e}{-iV(t) -\e}\right)^{\frac{1}{4}}.
$$
Notice that $K(z(t))$ is independent of the base point $a$ involved in the definition of the function $z(t)=z_a(t)$.
The branch of $K(z(t))$ is taken $e^{-i\frac{\pi}{4}}$ at each $t=t_k$ with 
the branch cut along a positive real axis on $\C_z$.
Here the vector-valued function $w_\pm(z;h) $ are determined as solutions of 
$$
\frac{d}{dz} w_\pm (z;h) = \begin{pmatrix}
0 & \frac{K'(z)}{K(z)}\\ \frac{K'(z)}{K(z)} & \mp \frac{2}{h}
\end{pmatrix}
w_\pm(z;h).
$$
Moreover, by identity \eqref{z_affine} and the following equality
\begin{equation}\label{sing_K'/K}
\frac{\frac{d}{dz}K(z(t))}{K(z(t))} = 
\frac{\alpha(t)\beta'(t) - \alpha'(t)\beta(t)}{4(\alpha(t)\beta(t))^{3/2}}\,,
\end{equation}
we see that $\frac{\frac{d}{dz}K(z(t))}{K(z(t))}$ and $w_\pm(z(t);h)$ are independent of $a$. 
The above equality \eqref{sing_K'/K} implies that 
when $\zeta$ is a simple turning point, 
the function $\frac{K'(z)}{K(z)}$ has a simple pole at $z=z(\zeta)$.\\

\noindent
Generally, even if the vector-valued symbols $w_\pm(z;h)$ are developed with respect to $h$ small enough, 
the series do not converge. 
The essential idea of \cite{GeGr88_01} (see also \cite{FuLaNe09_01}) is to introduce a resummation  
by using the following integral recurrence system on $\C_z$. 
More precisely, for any $b\in \widetilde {\mathcal S}_{\rm bdd}$, the vector-valued functions $w_\pm(z;h)=w_{\pm}(z,z(b);h)$ are of the form: 
\begin{equation}\label{resum1}
w_{\pm}(z,z(b);h) = \sum_{k\geq 0} w_{\pm,k}(z,z(b);h), 
 = \sum_{k\geq 0}\left(\begin{matrix}
w_{\pm,2k}(z,z(b);h)\\w_{\pm,2k-1}(z,z(b);h)
\end{matrix}\right),
\end{equation}
where the sequences $\big\{w_{\pm,k}(z,z(b);h)\big\}_{k\in\N}$ are defined by 
\begin{equation*}
\left\{
\begin{aligned}
w_{\pm,0}(z,z(b);h) &\equiv \, 1, \quad w_{\pm,-1}(z,z(b);h) \equiv \, 0,\\
\ w_{\pm,2k+1}(z,z(b);h) &=\int_{z(b)}^z 
e^{\pm\frac{2}{h}(\zeta-z)}\frac{K'(\zeta)}{K(\zeta)}
w_{\pm,2k}(\zeta,z(b);h)\,d\zeta  &&(k\geq 0),\\
w_{\pm,2k}(z,z(b);h) &=\int_{z(b)}^z 
\frac{K'(\zeta)}{K(\zeta)} w_{\pm,2k-1}(\zeta,z(b);h)\,d\zeta &&(k\geq 1).
\end{aligned}
\right.
\end{equation*}

Thanks to the above resummation, the vector-valued symbol expansions \eqref{resum1} converge
absolutely and uniformly in a neighborhood of $z(b)$ for $b\in \widetilde {\mathcal S}_{\rm bdd}$ (see, for example, \cite[Lemma 3.2]{FuLaNe09_01}). 
Hence, for any fixed $(a,b)\in {\mathcal S}_{\rm bdd}\times \widetilde {\mathcal S}_{\rm bdd}$, we can define 
the exact WKB solutions of type $\pm$ as follows:
\begin{align}\label{def_exactWKBsol}
\psi_{\pm}(t,a,b;h)
= \frac{1}{2}\begin{pmatrix}1&i\\i&1\end{pmatrix} e^{\pm \frac{z_a(t)}{h}} M_{\pm}(z(t))w_{\pm}(z(t),z(b);h), 
\end{align}
which are linearly independent exact solutions of \eqref{e1}.
Notice that $a\in {\mathcal S}_{\rm bdd}$ is the base point of the phase and $b\in \widetilde {\mathcal S}_{\rm bdd}$ is that of the symbol.\\

We conclude this subsection by recalling some results concerning the exact WKB solutions given by \eqref{def_exactWKBsol}.
In fact, the exact WKB method is based on two properties, which are the Wronskian formula 
between the exact WKB solutions of type $\pm$ and the asymptotic expansion with respect to $h$ of the symbol.

\begin{lemma}{\rm (\cite[\bf Proposition 2.2.2]{Wa06_01})} \label{wkblem1}
The Wronskian between any exact WKB solutions of type $\pm$ 
with the same base point of the phase satisfies:
\begin{equation}
\W [\psi_+(t,a,b_+;h), \psi_-(t,a,b_-;h)] = 2i \sum_{k\geq 0} w_{+,2k}(z(b_-),z(b_+);h),
\end{equation}
where $a\in {\mathcal S}_{\rm bdd}$ and $b_{\pm}\in \widetilde {\mathcal S}_{\rm bdd}$.
Here the Wronskian between $\C^2$-valued functions $\psi_1$ and $\psi_2$  
is defined by $\W[\psi_1,\psi_2] := {\rm det}\left( \psi_1 \psi_2\right)$.  
\end{lemma}

The proof of this lemma is based on a direct computation and 
the independence of the Wronskian with respect to the variable $t$ 
thanks to the trace-free matrix in \eqref{wkbModel}. 
The prefactor $2i$ is exactly the
$\det M_\pm$. \\

To state the next result, we introduce {\it canonical curves of type $\pm$} in $\widetilde {\mathcal S}_{\rm bdd}$ from a fixed point $b$ to $t$ 
along which $\pm \text{Re}\,z_a(t)$ increase strictly, for a fixed $a\in {\mathcal S}_{\rm bdd}$. 
The advantage of the integral recurrence system is to give not only an absolutely convergence 
but also $\C^2$-valued asymptotic sequences with respect to $h$ uniformly away from turning points. 
More precisely,

\begin{lemma}{\rm (\cite[\bf Proposition 2.3.1]{Wa06_01})}\label{wkblem2}
If there exist canonical curves of type $\pm$ from $b_\pm$ to $t$ denoted by $\gamma_\pm$,  
then the vector-valued symbols have the following asymptotic expansions:
\begin{equation}\label{symbolasymptotic}
w_{\pm}(z(t),z(b_\pm);h) = \begin{pmatrix}
1\\
0
\end{pmatrix}
\left( 1+ {\mathcal O}\left( \frac{h}{{\rm dist}(\gamma_\pm;\Lambda) }\right) \right)
\end{equation}
as $h$ tends to $0$, where ${\rm dist}(\gamma_\pm;\Lambda)$ stands for $\underset{t\in \gamma_\pm, \zeta\in \Lambda}{\inf}|z_{\zeta}(t)|$. 
\end{lemma}

This lemma can be proved by an integration by parts thanks to the exponential decaying along the canonical curve. \\

Combining Lemmas \ref{wkblem1} and \ref{wkblem2}, we obtain the asymptotic expansion of the Wronskian:
\begin{lemma}{\rm (\cite[\bf Proposition 2.4.1]{Wa06_01})}\label{wkblem3}
If there exists a canonical curve of type $+$ from $b_+$ to $b_-$ denoted by $\gamma$, 
the Wronskian between any exact WKB solutions of type $\pm$ with the same base point of the phase  
has the following asymptotic expansion, 
\begin{equation}\label{wkbWron}
\W [\psi_+(t,a,b_+;h), \psi_-(t,a,b_-;h)] =  
2i + {\mathcal O}\left( \frac{h}{{\rm dist}(\gamma;\Lambda) }\right)
\end{equation}
as $h$ tends to $0$, where ${\rm dist}(\gamma;\Lambda)=\underset{t\in \gamma, \zeta\in \Lambda}{\inf}|z_{\zeta}(t)|$. 
\end{lemma}

\begin{remark}[{``Adiabatic" v.s. ``Non-Adiabatic"}]\label{remWKB}

In order to derive the exponential decay of the transition probability (see Proposition \ref{thmWKB}),
we apply the above Wronskian formula to the local transfer matrix given by \eqref{T_kWron}),  
but we must find a canonical curve $\gamma$ passing between the two turning points $\zeta$ 
and $\overline{\zeta}$ which accumulate to the same crossing point as $\varepsilon$ tends to 0. 
In this case, one sees that ${\rm dist}(\gamma;\Lambda)$ is of order  $\Ord(\e^2)$. 
Hence the WKB method works only under the regime ${{\scriptstyle\frac{h}{\e^2}}} \to 0$, called ``adiabatic" regime.

\smallskip\noindent
When ${\scriptstyle\frac{\e^2}{h}} \to 0$, the above lemma is obsolete at the crossing point. 
But we can control the behavior of the error in \eqref{wkbWron} far from this point, for example
outside an $\mathcal{O}(\sqrt{h})$-neighborhood of it. We say this regime ``non-adiabatic".
\end{remark}

\subsection{{ Representation of the scattering matrix}}\label{S_matrix_section}

\begin{figure}
\begin{center}
\scalebox{0.35}[0.35]{
\includegraphics{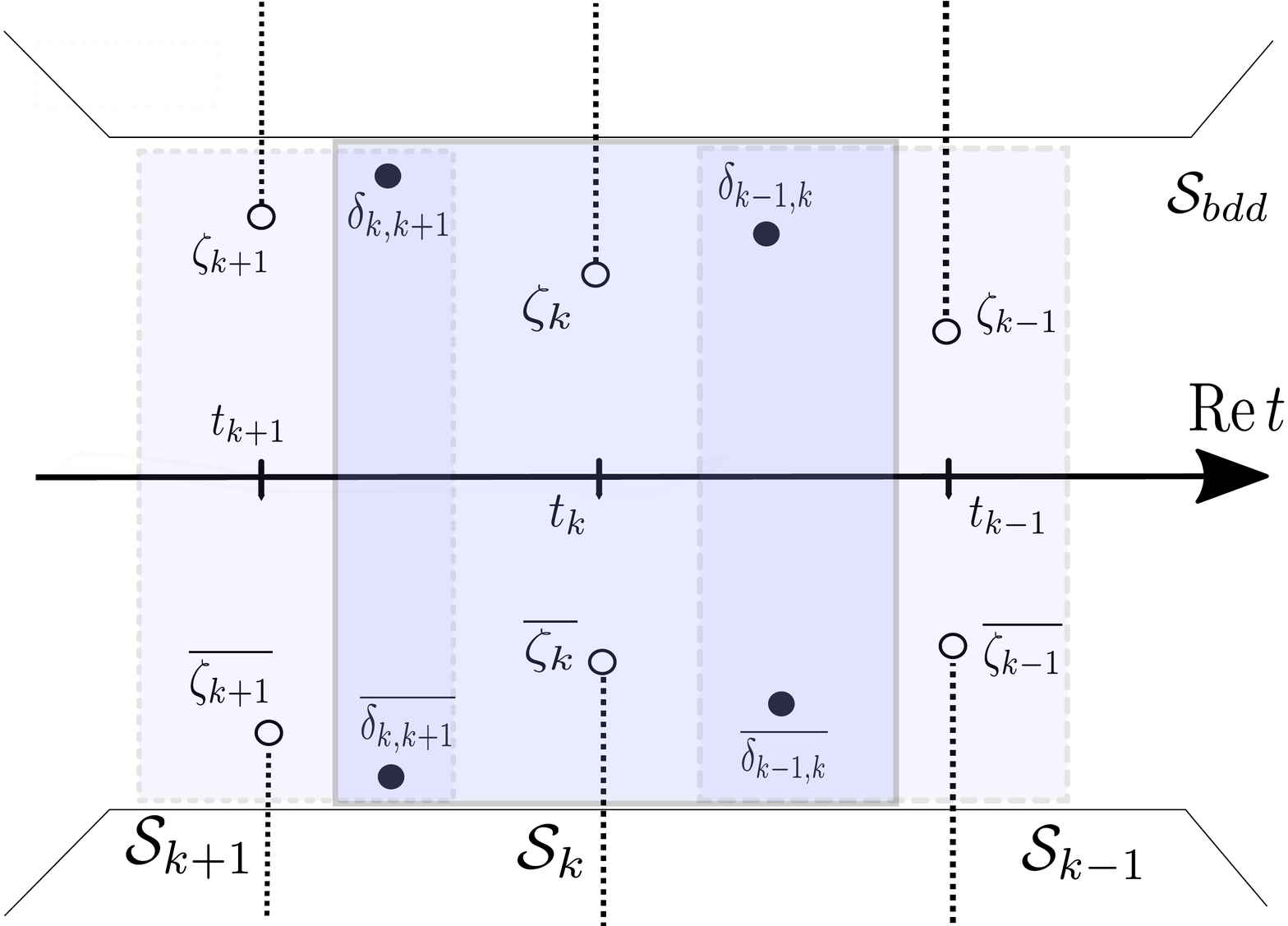}
}
\caption{Picture of ${\mathcal S}_k$}
\label{S_box}
\end{center}
\end{figure}

In this subsection we give the proof of Proposition \ref{S-represent}.
When we construct exact WKB solutions globally 
for a sake of expressing the scattering matrix, 
it is difficult to deal with various turning points, so that 
we treat only two turning points near each vanishing point of $V(t)$, without loss of generality.
We put $\scriptstyle d_1 = \frac{t_1 - t_2}{2}$, $\scriptstyle d_n = \frac{t_{n-1} - t_n}{2}$ and 
$d_k =\frac{1}{2} \max\{ t_k - t_{k+1},\ t_{k-1} - t_k \}$ for $k= 2,\ldots, n-1$. 
Let ${\mathcal S}_k \subset {\mathcal S}_{\rm bdd}$ be 
a simply connected small box in ${\mathcal S}_{\rm bdd}$ 
including only one vanishing point $t_k$, given by 
\begin{equation}\label{Sbox_k}
{\mathcal S}_k := \{ t\in {\mathcal S}_{\rm bdd}\ ; \ 
|\text{Re}\,t - t_k|< d_k +\rho \}\quad (k=1,2,\ldots, n),
\end{equation}
where $\rho$ is a suitable small constant. 
We see that ${\mathcal S}_k \cap {\mathcal S}_{k+1} \neq \emptyset$ for each $k= 1,\ldots, n-1$ 
and we put there a symbol base point 
$$\delta_{k,k+1} := \frac{t_k + t_{k+1}}{2} + i c_{k} \in {\mathcal S}_k \cap {\mathcal S}_{k+1}$$ 
and its complex conjugate (see Figure \ref{S_box}). 
Here $0<c_k<c$ where the constant $c$ is involved in the definition of 
${\mathcal S}_{\rm bdd}$.


In each ${\mathcal S}_k$ $(k=1,\ldots, n)$, 
we introduce the intermediate WKB solutions, which consist of the bases in ${\mathcal S}_k$,   
\begin{equation}\label{4wkb_S_k} 
\begin{aligned}
\psi_+(t,\zeta_k,\delta_{k-1,k};h)
&= \exp \left[+\frac{z_{\zeta_k}(t)}{h}\right]
M_+(z(t))w_+(z(t),z(\delta_{k-1,k});h) &&:= \psi_{+,k}^r,\\
\psi_-^{\vphantom{b}}(t,\overline{\zeta_k},\overline{\delta_{k-1,k}};h)
&= \exp \left[-\frac{z_{\overline{\zeta_k}}(t)}{h}\right]  
M_-(z(t))w_-(z(t),z(\overline{\delta_{k-1,k}});h) &&:= \psi_{-,k}^r,\\
\psi_+(t,\zeta_k,\delta_{k,k+1};h) 
&= \exp \left[+\frac{z_{\zeta_k}(t)}{h}\right]
M_+(z(t))w_+(z(t),z(\delta_{k,k+1});h) &&:= \psi_{+,k}^l,\\
\psi_-^{\vphantom{b}}(t,\overline{\zeta_k},\overline{\delta_{k,k+1}};h)
&= \exp \left[-\frac{z_{\overline{\zeta_k}}(t)}{h}\right]
M_-(z(t))w_-(z(t),z(\overline{\delta_{k,k+1}});h) &&:= \psi_{-,k}^l.
\end{aligned}
\end{equation}

Remark that each exact WKB solution has a valid asymptotic expansion for $h$ small enough 
in the direction from its symbol base point toward the vanishing point $t_k$ in ${\mathcal S}_k$ 
thanks to Lemma \ref{wkblem2}.\\

As mentioned in Subsection \ref{def_ScatteringM}, there exist two kind of the transfer matrices. 
One of them is a change of bases with respect to the base points of the symbol function 
denoted by $T_k(\e,h)$, that is, it transfers from right side to left side over the crossing point in ${\mathcal S}_k$. 
The other is one with respect to the base points of the phase function denoted by $T_{k,k+1}(\e,h)$, 
that is, it transfers on the intersection between ${\mathcal S}_k$ and ${\mathcal S}_{k+1}$. 
In fact, they are written by 
\begin{align}\label{defT_k}
\left(\psi_{+,k}^l \ \psi_{-,k}^l \right) 
= \left(\psi_{+,k}^r\ \psi_{-,k}^r\right)T_{k},\quad 
\left(\psi_{+,k+1}^r \ \psi_{-,k+1}^r \right) 
= \left(\psi_{+,k}^l\ \psi_{-,k}^l\right)
T_{k,k+1}.
\end{align}
The former transfer matrix $T_k(\e,h)$ is our main target, which is given by   
\begin{equation}\label{T_kWron}
T_k(\e,h) = \frac{1}{\W[\psi_{+,k}^r, \psi_{-,k}^r]} 
\begin{pmatrix}
\W[\psi_{+,k}^l, \psi_{-,k}^r] & \W[\psi_{-,k}^l, \psi_{-,k}^r]\\
\W[\psi_{+,k}^r, \psi_{+,k}^l] & \W[\psi_{+,k}^r, \psi_{-,k}^l]
\end{pmatrix}.
\end{equation}
The asymptotic behaviors of these Wronskians can be computed by Lemma \ref{wkblem3} 
under the regime ${\scriptstyle \frac{h}{\e^2}} \to 0$ (see Lemma \ref{adiabatic_T_k}). 
On the other hand, the asymptotic behaviors of them under the regime ${\scriptstyle \frac{\e^2}{h}} \to 0$ 
must be investigated more carefully (see Remark \ref{remWKB}).

\smallskip\noindent
From \eqref{z_affine}, the latter transfer matrix $T_{k,k+1}(\e,h)$ is the diagonal one, 
whose diagonal elements are complex conjugate each other. 
Put
\begin{equation}\label{a_k}
a_k(\e,h) = e^{\frac{i}{2h}\left(A_k(\e) - A_{k+1}(\e) + R_{k}(\e)\right)}
\end{equation}
for $k = 1,2, \ldots, n-1$, where $A_k(\e)$ (resp. $R_k(\e)$) is given by \eqref{Actionk} (resp. \eqref{Actionjk}).   
Then, for $k = 1,2, \ldots, n-1$,
\begin{align}\label{T_kk}
T_{k,k+1}(\e,h) 
&= \left(
\begin{array}{cc}
a_k(\e,h) & 0\\
0 & \overline{a_k(\e,h)}
\end{array}
\right).
\end{align}

\noindent
In addition, the transfer matrices between Jost solutions and exact WKB solutions at $\pm\infty$ 
must be treated separately. For a fixed $R>0$ we denote an unbounded simply connected domain by 
${\mathcal S}_R = {\mathcal S}\cap \{ t\in\C\, ;\ |{\rm Re}\, t | > R\}$. 
From \ref{A2}, we can find a suitable constant $R>0$ such that ${\mathcal S}_R \cap \Lambda = \emptyset$.
Recall that $\Lambda$ is the set of turning points.
In ${\mathcal S}_R$ we can construct the exact WKB solutions of \eqref{wkbModel} corresponding to Jost solutions as 
\begin{align*}
\phi_\pm^\star (t;h) = e^{\pm \frac{z_{\star}(t)}{h}} M_\pm(z(t)) w_\pm^\star (z(t);h), 
\end{align*}
where the index $\star\in \{r,l\}$ stands for a direction corresponding to either $+\infty$ or $-\infty$, 
and a modified phase function $z_{\star}(t)$ 
is given by
\begin{align*}
z_{\star}(t) &= i \int_{\pm \infty}^t \left(\sqrt{V^2(s) + \e^2} - \lambda_\star \right) ds + i\lambda_\star\, t,
\end{align*}
with $\lambda_\star = \sqrt{E_\star^2 + \e^2}$ 
and a modified symbol functions $w_\pm^{\star}(z(t))$ are given by the same integral recurrence system 
described in Subsection \ref{constWKB},
but the integral paths taken from $\infty e^{\pm i\theta_1}$ to $t\in {\mathcal S}_R$
for $\theta_1\in (0,\theta_0)$ with $\theta_0$ given by \ref{A1}.
Remark that the modified phase functions are convergent thanks to \ref{A2} and 
the resummation based on the modified integral paths works also thanks to \ref{A1}. 

\begin{lemma}{\rm (\cite[\bf Proposition 3.1.1]{Wa06_01})} 
We obtain the relation between Jost solutions and the corresponding WKB solutions:
\begin{align*}
&J_+^\star(t) = -\frac{1}{2}\begin{pmatrix}1&i\\i&1\end{pmatrix} \phi_+^\star(t),
\quad J_-^\star(t) = -\frac{i}{2}\begin{pmatrix}1&i\\i&1\end{pmatrix} \phi_-^\star(t),\qquad (\star\in \{r,l\}).
\end{align*}
\end{lemma}
One sees that the computations of the asymptotic behaviors of $M_\pm(z(t))$ are essentially 
same as in Subsection \ref{wkbsec}  and about those of $w_\pm^{\star}(z(t);h)$ 
one can consult with \cite{Ra96_01} (see also \cite[Lemma 3.2]{Gr87_01}), in fact 
$w_\pm^{\star}(z(t);h) \to { }^t (1, 0)$ as ${\rm Re}\, t \to \pm \infty$. 
From this lemma, when we denote by $T_r(\e,h)$ (resp. $T_l(\e,h)$) 
the transfer matrices from ${\mathcal S}_R$ to ${\mathcal S}_1$ (resp. ${\mathcal S}_n$)  as 
\begin{align*}
(J_+^r\ J_-^r) = (\psi_{+,1}^r \ \psi_{-,1}^r)T_r,\quad (J_+^l\ J_-^l) = (\psi_{+,n}^l \ \psi_{-,n}^l)T_l,
\end{align*}
and do the action integrals corresponding to $\pm\infty$ by
\begin{align}\label{action_infty}
A_{\star}(\e) &= 2\int_{t_\star}^{\pm\infty}
{\left(\sqrt{V(t)^2 + \e^2} - \sqrt{E_\star^2+\e^2}\right)}\,dt,
\end{align}
with $t_r=t_1$ and $t_l = t_n$, the transfer matrices $T_r(\e,h)$ and $T_l(\e,h)$ are diagonal.
Actually, they are given by 
\begin{align}\label{r-inf}
T_{\star}(\e,h) &= \left(
\begin{array}{cc}
-a_\star (\e,h) & 0\\
0 & \overline{ia_\star(\e,h)}
\end{array}
\right)\Bigl(1+f_\star(h)\Bigr),
\end{align}
where each error $f_\star(h)$ $(\star\in\{r,l\})$ is ${\mathcal O}(h)$ as $h$ tends to $0$ 
uniformly with respect to small $\e$ and 
\begin{equation}\label{a_r_l}
a_r (\e,h) = e^{\frac{i}{2h}\left(A_1(\e)-A_{r}(\e)+2\lambda_rt_1\right)},\qquad 
a_l(\e,h) = e^{\frac{i}{2h}\left(A_n(\e)-A_{l}(\e)+2\lambda_lt_n\right)}.
\end{equation}

Summing up, by using all kinds of the transfer matrices, we have a representation of the scattering matrix as 
we state in Proposition \ref{S-represent}. 

\begin{remark}\label{same_config}
The asymptotic behaviors of the scattering matrix $S(\e,h)$ are essentially given by the asymptotic expansions of
the local transfer matrices $(T_k)_{1\leq k \leq n}$. So, the shape of the function $V(t)$ near its vanishing points
is crucial. The assumption \ref{A3} implies that we have the same geometrical configuration in each ${\mathcal S}_k$ 
for $k= 1,2,\ldots, n$. Hence, without loss of generality,  we may assume that $t_k = 0$ 
and $V(0) = 1$, that is, $V(t) = t + {\mathcal O}(t^2)$ in an $h$-independent neighborhood of 0. 
\end{remark}

\subsection{{ Proof of Proposition \ref{AsymWKB}}}\label{proof_WKB} 

For the proof of Proposition \ref{AsymWKB}, it is enough to compute the leading term
of the exact WKB solutions $\widetilde \psi_{\pm}^\star$ with $\star\in\{r,l\}$.
Remembering that $V(t)$ is bounded and real-valued in $I^\star(h)$, where $I^\star(h)$ is given by \eqref{hrealdomain}, 
we first study $K(z(t)) = (\kappa(t))^{\frac{1}{4}}$ and see that 
\begin{align*}
\kappa(t) = \frac{-iV(t) + \e}{-iV(t) - \e} 
&= \frac{V(t)^2 - \e^2}{V(t)^2 + \e^2} + i \frac{2\e V(t)}{V(t)^2 + \e^2}\quad {\rm and}
\quad |\kappa(t)| = 1.
\end{align*} 
Recall that, under our branch of $K$, the function $\kappa(t)$ tends to $e^{-\pi i}$ as $t$ goes to $0$. 
We express $\kappa(t)$ and $K(z(t))$ as follows:
\begin{align*}
\kappa(t) &= \left\{
\begin{aligned}
&e^{i(\theta(t;\e) - 2\pi)} && \text{for} \ t>0,\\
&e^{-i\theta(t;\e)} && \text{for} \ t<0,
\end{aligned}
\right.
\quad {\rm and}\quad 
K(z(t)) = \left\{
\begin{aligned}
 -i e^{i\frac{\theta(t;\e)}{4}} &\quad \text{for} \ t>0,\\
e^{-i\frac{\theta(t;\e)}{4}} &\quad \text{for} \ t<0,
\end{aligned}
\right.
\end{align*} 
where $\theta(t;\e)$ satisfies $\tan \theta(t;\e) = \frac{2\e |V(t)|}{|V(t)|^2-\e^2}$ for $0< \theta(t;\e) <\frac{\pi}{2}$. 
In fact, we know 
$$
\theta(t;\e) = \tan^{-1} \frac{2\e |V(t)|}{|V(t)|^2-\e^2} = \frac{2\e}{|V(t)|} + \Ord\left(\left(\frac{\e}{\sqrt{h}}\right)^3\right).
$$
Hence we have 
\begin{align*}
 & K(z(t))^{-1} + K(z(t))\\
 =\ & \left\{
\begin{aligned}
&2\sin \frac{\theta(t;\e)}{4} = \frac{\e}{t}(1 + \Ord(\sqrt{h})) + \Ord\left(\left(\frac{\e}{ \sqrt{h}}\right)^{3}\right)
 &&\qquad \text{for} \ t>0,\\
&2\cos \frac{\theta(t;\e)}{4} = 2 + \Ord\left(\frac{\e^2}{ h}\right) &&\qquad \text{for} \ t<0,
\end{aligned}
\right.\\[7pt]
 & K(z(t))^{-1} - K(z(t)) \\
=\ & \left\{
\begin{aligned}
&2i\cos \frac{\theta(t;\e)}{4} = 2i + \Ord\left(\frac{\e^2}{h}\right) &&\quad \text{for} \ t>0,\\
&2i\sin \frac{\theta(t;\e)}{4} = -i\frac{\e}{t}(1 + \Ord(\sqrt{h})) + \Ord\left(\left(\frac{\e}{\sqrt{h}}\right)^{3}\right)
  &&\quad \text{for} \ t<0,
\end{aligned}
\right.
\end{align*}
as $\e$ and $h$ go to $0$ and $\frac{\e^2}{h}$ tends to $0$ uniformly in $I^\star(h)$.
These last asymptotic expansions give us the behaviors of the leading terms involved in the vector-valued symbols of the exact WKB solutions.\\

Next, we give the asymptotic behaviors of the phase functions for $t \in I^r(h)$, 
by the same way we have them in $I^l(h)$.

\noindent
We decompose the phase functions as follows, involving the crossing point 0,   
\begin{equation}\label{phase1}
\int_{\zeta_\pm}^t \sqrt{V(s)^2+\e^2}\,ds
 = \int_0^t \sqrt{V(s)^2+\e^2}\,ds - \int_0^{\zeta_\pm} \sqrt{V(s)^2+\e^2}\,ds.
\end{equation}
The second integral of \eqref{phase1} is the action integral $A(\zeta_\pm)$, 
which is $\Ord(\e^2)$.  
The first of \eqref{phase1} is real-valued and moreover can be decomposed with some constant $c>0$ as  
\begin{equation}\label{phase2}
\int_0^t \sqrt{V(s)^2+\e^2}\,ds = \int_0^{c \e} \sqrt{V(s)^2+\e^2}\,ds + \int_{c \e}^t \sqrt{V(s)^2+\e^2}\,ds.
\end{equation}
The first integral of \eqref{phase2} is also $\Ord(\e^2)$. 
We put 
\begin{align*}
F(s; \e) &:= \sqrt{V(s)^2 + \e^2} - V(s) - \frac{\e^2}{2V(s)} = \frac{-\e^4}{2V(s)(\sqrt{V(s)^2 + \e^2}+V(s))^2},\\
G(s) &:= \frac{1}{V(s)} - \frac{1}{s} = \frac{-V_1(s)}{s(1+V_1(s))},
\end{align*}
where $V_1$ is a holomorphic  function satisfying $V(s) = s(1+V_1(s))$ such that $V_1(s) = \Ord(s)$. 
Then we can estimate, by using the fact that $|1+ V_1(s)|^{-1} \leq 2$ along the integral path on $I^r(h)$
 for $h$ small enough, the integrals of the absolute value of these functions as 
\begin{align*}
\int_{c\e}^t |F(s; \e)| ds &= \Ord(\e^2),\qquad 
\int_{c\e}^t |G(s)| ds = \Ord(\sqrt{h}).
\end{align*}
These estimates imply that
\begin{align*}
 &\int_{c\e}^t \sqrt{V(s)^2 + \e^2} ds\\ 
= \ &\int_{c\e}^t V(s)ds + \frac{\e^2}{2} \int_{c\e}^t \frac{ds}{V(s)} + \Ord(\e^2)\\
= \ &\left(\int_{0}^t V(s)ds + \Ord(\e^2)\right) + \frac{\e^2}{2} \left(\int_{c\e}^t \frac{ds}{s} + \Ord(\sqrt{h})\right) + \Ord(\e^2)\\
= \ &\int_{0}^t V(s)ds + \frac{\e^2}{2}\log t - \frac{\e^2}{2}\log \e + \Ord(\e^2).
\end{align*}
Hence we have 
\begin{align*}
e^{\textstyle\frac{i}{h}\int_{\zeta_+}^t\sqrt{V(s)^2 + \e^2}ds} &= e^{\textstyle \frac{i}{h}\int_{0}^t V(s)ds} t^{\frac{i\e^2}{2h}} e^{-\frac{i\e^2}{2h}\log \e} 
\left( 1 + \Ord\left( \frac{\e^2}{h} \right)\right).
\end{align*}

Finally, in the case where $(\e,h)$ goes to $(0,0)$ and $\frac{\e^2}{h}$ tends to $0$, 
we combine the asymptotic behaviors of each part  
in the intervals $I^\star(h)$, and then we obtain Proposition \ref{AsymWKB}.

\section{{\sl Branching-model and its applications}}\label{BM_application}

\subsection{{ Solutions of the branching model}}\label{Sol_BM}

The branching model:
\begin{align}\label{BM_app}
Q\phi(y) = \left(
\begin{matrix}
y & \frac{\mu}{\sqrt{2}}\\
\frac{\mu}{\sqrt{2}} & \frac{\mu}{i}\frac{d}{dy}
\end{matrix}
\right)
\phi(y) = 0
\end{align}
with a small parameter $\mu >0$ has two solutions of the forms:
\begin{equation}\label{onebasis}
\begin{aligned}
\phi^{\vdash}(y) &= Y(y)y^{\frac{i}{2}\mu}\left(
\begin{matrix}
-\frac{\mu}{\sqrt{2}y} \\ 1
\end{matrix}
\right),\\
\phi^{\dashv}(y) &= Y(-y)|y|^{\frac{i}{2}\mu}\left(
\begin{matrix}
-\frac{\mu}{\sqrt{2}y} \\ 1
\end{matrix}
\right),
\end{aligned}
\end{equation}
where $Y(y)$ is the Heaviside function. 
The properties of these distributions can be found in \cite{GuCh62_01}. 
Notice that this equation \eqref{BM_app} is treated in Subsection \ref{MicroReduction}. 
Here we give properties of the solutions \eqref{onebasis}. 

The differential operator $Q$ commutes with the operator 
$\left(\begin{matrix}
0&1\\1&0
\end{matrix}
\right){\mathcal C} {\mathcal F}_\mu$, 
where ${\mathcal C}$ is a complex conjugate operator, that is ${\mathcal C}\phi(y) = \overline{\phi(y)}$, and 
${\mathcal F}_\mu$ is a semi-classical Fourier transform:
\begin{equation}\label{F_mu}
{\mathcal F}_\mu[u](\xi) = \frac{1}{\sqrt{2\pi \mu}}
\int_{\R} e^{-\frac{i}{\mu}x\xi} u(x) dx.
\end{equation} 
Then the functions $\phi^{\bot}(y)$ and $\phi^{\top}(y)$ given by 
\begin{align}\label{otherbasis}
\phi^{\bot}(y) &= \left(
\begin{matrix}
0&1\\1&0
\end{matrix}
\right){\mathcal C} {\mathcal F}_\mu \phi^{\vdash}(y), \qquad 
\phi^{\top}(y) = \left(
\begin{matrix}
0&1\\1&0
\end{matrix}
\right){\mathcal C} {\mathcal F}_\mu \phi^{\dashv}(y)
\end{align}
are also solutions of \eqref{BM_app}. 
Computing $\phi^{\bot}(y)$, $\phi^{\top}(y)$ by using the property 
${\mathcal C} {\mathcal F}_\mu = {\mathcal F}_\mu^{-1} {\mathcal C}$, 
we obtain the relation between the pairs $(\phi^{\vdash}, \phi^{\dashv})$ and $(\phi^{\bot}, \phi^{\top})$. 
\begin{proposition}\label{ChangeBasisNFsol} 
Let $R$ be a $2\times 2$ matrix such that
$(\phi^\bot \phi^\top) = (\phi^\vdash \phi^\dashv)R$. 
Then $R$ is of the form: 
$R = \left(
\begin{matrix}
p & -q\\
q & -p
\end{matrix}
\right)$ 
with 
\begin{equation}\label{pqgamma}
p = \gamma e^{\frac{\pi \mu}{4}}, \qquad 
q = \gamma e^{-\frac{\pi \mu}{4}},\qquad 
\gamma = 
\frac{1}
{i\sqrt{\pi\mu}}  \mu^{-\frac{i\mu}{2}} \Gamma\left(1-\frac{i\mu}{2}\right).
\end{equation}
\end{proposition}

\noindent
{\sl Proof of Proposition \ref{ChangeBasisNFsol}}:
The direct computations of $\phi^{\bot}(y)$ and $\phi^{\top}(y)$ under the definition \eqref{otherbasis} 
give us the entries of $\phi^{\bot}(y)$ and $\phi^{\top}(y)$ 
expressed by those of $\phi^{\vdash}(y) = \, ^t(\phi_1^{\vdash}, \phi_2^{\vdash})$ and $\phi^{\dashv}(y) = \, ^t(\phi_1^{\dashv}, \phi_2^{\dashv})$ as 
$$
\left(
\begin{matrix}
\phi_1^{\star}\\ \phi_2^{\star}
\end{matrix}
\right) = \left(
\begin{matrix}
0&1\\1&0
\end{matrix}
\right)\left(
\begin{matrix}
{\mathcal C}{\mathcal F}_\mu[\phi_1^{\ast}]\\[5pt]
{\mathcal C}{\mathcal F}_\mu[\phi_2^{\ast}]
\end{matrix}
\right)
 = \left(
\begin{matrix}
{\mathcal C}{\mathcal F}_\mu[\phi_2^{\ast}]\\[5pt]
{\mathcal C}{\mathcal F}_\mu[\phi_1^{\ast}]
\end{matrix}
\right),
$$
where the pair of the indexes $(\ast,\star) \in \{ (\vdash, \bot), (\dashv, \top) \}$. 
For example, the first entries of $\phi^{\bot}(y)$ are expressed by 
$$
\phi_1^{\bot}(y) = {\mathcal C}{\mathcal F}_\mu[\phi_2^{\vdash}](y) 
= {\mathcal F}_\mu^{-1}[{\mathcal C}\phi_2^{\vdash}](y) 
= \frac{1}{\sqrt{2\pi \mu}}\int_{\R} e^{\frac{i}{\mu}y\eta} Y(\eta)
\eta^{-\frac{i}{2}\mu} d\eta.
$$

We demonstrate the computation concerning only $\phi^\bot_1$.
\begin{align*}
\phi^\bot_1 (y) &= 
\frac{1}{\sqrt{2\pi \mu}}\int_{\R} e^{\frac{i}{\mu}y \eta} Y(\eta)
|\eta|^{-\frac{i}{2}\mu} d\eta 
= \frac{1}{\sqrt{2\pi \mu}}\int_{0}^{\infty} e^{\frac{i}{\mu}y \eta} 
\eta^{-\frac{i}{2}\mu} d\eta.
\intertext{In order to reduce this integral to the Gamma function, 
we treat it separately as a positive part and a negative part with respect to $y$.}
\phi_1^\bot (y) &= 
\frac{Y(y)}{\sqrt{2\pi \mu}}\int_{0}^{\infty} e^{\frac{i}{\mu}y \eta} 
\eta^{-\frac{i}{2}\mu} d\eta +
\frac{Y(-y)}{\sqrt{2\pi \mu}}\int_{0}^{\infty} e^{\frac{i}{\mu}y \eta} 
\eta^{-\frac{i}{2}\mu} d\eta.
\end{align*}
For the first integral, by the change of variable $\frac{i}{\mu}y\eta = -z$ and by the Cauchy integral theorem, we have
\begin{align*}
\frac{1}{\sqrt{2\pi \mu}}\int_{0}^{\infty} e^{\frac{i}{\mu}y \eta} 
\eta^{-\frac{i}{2}\mu} d\eta
 &= \frac{i}{\sqrt{2\pi}} \mu^{\frac{1}{2}-\frac{i}{2}\mu}
e^{\frac{\pi}{4}\mu} \Gamma \left(1-\frac{i}{2}\mu\right)
y^{\frac{i}{2}\mu -1}.
\intertext{Recalling the form of the first entry of $\phi^\vdash$, we see} 
\frac{1}{\sqrt{2\pi \mu}}\int_{0}^{\infty} e^{\frac{i}{\mu}y \eta} 
\eta^{-\frac{i}{2}\mu} d\eta
 &= \frac{-i}{\sqrt{\pi}\mu} \mu^{\frac{1}{2}-\frac{i}{2}\mu}
e^{\frac{\pi}{4}\mu} \Gamma \left(1-\frac{i}{2}\mu\right) \phi_1^\vdash(y).
\end{align*}
Similarly we compute the second one with the change of variable 
$\frac{i}{\mu}(-y)\eta = z$ as
\begin{align*}
\frac{1}{\sqrt{2\pi \mu}}\int_{0}^{\infty} e^{-\frac{i}{\mu}(-y) \eta} 
\eta^{-\frac{i}{2}\mu} d\eta 
 &= 
\frac{i}{\sqrt{\pi}\mu} \mu^{\frac{1}{2}-\frac{i}{2}\mu}
e^{-\frac{\pi}{4}\mu} \Gamma \left(1-\frac{i}{2}\mu\right) \phi_1^\dashv(y).
\end{align*}
Therefore we obtain
\begin{align*}
\phi_1^\bot(y) &= \frac{1}{i\sqrt{\pi\mu}} \mu^{-\frac{i}{2}\mu}
 \Gamma \left(1-\frac{i}{2}\mu\right)
\left(
e^{\frac{\pi}{4}\mu}  \phi_1^\vdash(y) + e^{-\frac{\pi}{4}\mu} \phi_1^\dashv(y)
\right).
\intertext{By similar computations, we have the followings:}
\phi_2^\bot(y) &= \frac{1}{i\sqrt{\pi\mu}} \mu^{-\frac{i}{2}\mu}
 \Gamma \left(1-\frac{i}{2}\mu\right)
\left(
e^{\frac{\pi}{4}\mu}  \phi_2^\vdash(y) + e^{-\frac{\pi}{4}\mu} \phi_2^\dashv(y)
\right),\\
\phi_1^\top(y) &= \frac{1}{i\sqrt{\pi\mu}} \mu^{-\frac{i}{2}\mu}
 \Gamma \left(1-\frac{i}{2}\mu\right)
\left(
-e^{-\frac{\pi}{4}\mu}  \phi_1^\vdash(y) - e^{\frac{\pi}{4}\mu} \phi_1^\dashv(y)
\right),\\
\phi_2^\top(y) &= \frac{1}{i\sqrt{\pi\mu}} \mu^{-\frac{i}{2}\mu}
 \Gamma \left(1-\frac{i}{2}\mu\right)
\left(
-e^{-\frac{\pi}{4}\mu}  \phi_2^\vdash(y) - e^{\frac{\pi}{4}\mu} \phi_2^\dashv(y)
\right).
\end{align*}
Hence we get the relation betweens $(\phi^\vdash, \phi^\dashv)$ and $(\phi^\bot, \phi^\top)$.
$$
\left( \phi^\bot \phi^\top \right)
= 
\frac{1}
{i\sqrt{\pi\mu}} 
\mu^{-\frac{i}{2}\mu}  
\Gamma \left(1-\frac{i}{2}\mu\right)
\left(
\phi^\vdash \phi^\dashv
\right)
\left(
\begin{matrix}
{\displaystyle e^{\frac{\pi}{4}\mu} } &
{\displaystyle -e^{-\frac{\pi}{4}\mu} } \\
{\displaystyle e^{-\frac{\pi}{4}\mu} } &
{\displaystyle -e^{\frac{\pi}{4}\mu} }
\end{matrix}
\right).
$$

\hfill$\Box$

From the reflection property of the Gamma function:
$$
\Gamma(z)\Gamma(1-z) = \frac{\pi}{\sin \pi z} 
= \frac{2\pi i}{e^{i\pi z} - e^{-i\pi z}}\qquad (z\in\C\setminus \Z)
$$
and thanks to $\frac{q}{p}\in \R$,
we get the properties of the constants $\gamma$, $p$ and $q$ as follows:
\begin{align}\label{pq}
|\gamma|^2 = \left(e^{\frac{\pi}{2}\mu} - e^{-\frac{\pi}{2}\mu}\right)^{-1},\quad 
|p|^2 - |q|^2 = 1, \quad \frac{p^2 - q^2}{p} = \frac{1}{\bar{p}}.
\end{align}

\subsection{{ Asymptotic expansions of pull-back solutions of the branching model}}

In this subsection, we give the asymptotic behaviors of the images of Fourier integral operator $U_{\frac{\pi}{4}}$ 
of the solutions of the branching model $Q\phi = 0$, which are studied in Appendix \ref{Sol_BM}. 
We put $I_c(\mu) := \{ x\in\R ;\ c \sqrt{\mu} \leq |x| \leq 2c \sqrt{\mu} \}$, for some constant $c>0$,
where the notation $I_c(\mu)$ is introduced in \eqref{coer_interval}.   
\begin{proposition}\label{Asymlocal} 
There exists $\mu_0 >0$ small enough such that for any $\mu \in (0, \mu_0]$ we obtain  
\begin{align*}
U_{\frac{\pi}{4}} [\phi^\vdash](x) &= e^{-\frac{\pi}{8}i} 2^{\frac{1}{4}} e^{\frac{ix^2}{2\mu}} x^{\frac{i}{2}\mu} 
\begin{pmatrix}
-\frac{\mu}{2x} \\ 1
\end{pmatrix} 
\left( 1 + {\mathcal E}^\vdash (x; \mu) \right)&&(x\in I_c(\mu)\cap \R_+),\\
U_{\frac{\pi}{4}} [\phi^\dashv](x) &= e^{-\frac{\pi}{8}i} 2^{\frac{1}{4}} e^{\frac{ix^2}{2\mu}} (-x)^{\frac{i}{2}\mu} 
\begin{pmatrix}
-\frac{\mu}{2x} \\ 1
\end{pmatrix} 
\left( 1 + {\mathcal E}^\dashv (x; \mu) \right) &&(x\in I_c(\mu)\cap\R_-),\\
U_{\frac{\pi}{4}} [\phi^\bot](x) &= e^{\frac{\pi}{8}i} 2^{\frac{1}{4}} e^{-\frac{ix^2}{2\mu}} (-x)^{-\frac{i}{2}\mu} 
\begin{pmatrix}
1 \\ \frac{\mu}{2x}
\end{pmatrix} 
\left( 1 + {\mathcal E}^\bot (x; \mu) \right) &&(x\in I_c(\mu)\cap \R_-),\\
U_{\frac{\pi}{4}} [\phi^\top](x) &= e^{\frac{\pi}{8}i} 2^{\frac{1}{4}} e^{-\frac{ix^2}{2\mu}} x^{-\frac{i}{2}\mu} 
\begin{pmatrix}
1 \\ \frac{\mu}{2x}
\end{pmatrix} 
\left( 1 + {\mathcal E}^\top (x; \mu) \right) &&(x\in I_c(\mu)\cap \R_+),
\end{align*}
where each error ${\mathcal E}^{\ast} (x;\mu)$ is a function satisfying   
${\mathcal E}^{\ast} (x;\mu) = \Ord (\mu^2 |x|^{-2}) = \Ord(\mu)$ uniformly on $I_c(\mu)$ 
with $\ast \in \{ \vdash, \dashv, \bot, \top\}$.
\end{proposition}

\noindent
{\sl Proof of Proposition \ref{Asymlocal}}: 
The proof is the direct calculation by using the stationary phase method. 
We show the calculation of the asymptotic behavior of $U_{\frac{\pi}{4}}[\phi^\vdash](x)$ 
when $\mu$ tends to $0$ for $x \in I_c(\mu)\cap\R_+$. 
From the definition of the Fourier integral operator $U_{\frac{\pi}{4}}$, 
we compute 
\begin{align*}
U_{\frac{\pi}{4}}[\phi_2^\vdash(y)](x) &= \frac{e^{\frac{\pi i}{8}}2^{\frac{1}{4}}}
{\sqrt{2\pi \mu}} e^{-\frac{ix^2}{2\mu}} \int_{0}^{\infty} 
e^{\frac{i}{\mu}f(x,y)}
y^{\frac{i\mu}{2}}  dy,
\end{align*}
where the phase function $f(x,y)$ is of the form: $f(x,y) = \sqrt{2}xy - \frac{y^2}{2}$.
The first and second derivatives of $f(x,y)$ with respect to $y$ are  
$f_y (x,y) = \sqrt{2}x - y$ and $f_{yy} (x,y) = -1$. 
One sees that the stationary point $y = \sqrt{2}x = \Ord(\sqrt{\mu})$ 
lies on the finite integral path for $x \in I_c(\mu)\cap\R_+$.  
Applying the stationary phase method to $U_{\frac{\pi}{4}}[\phi_2^\vdash](x)$, we obtain 
\begin{align*}
U_{\frac{\pi}{4}}[\phi_2^\vdash](x) 
 &= e^{-\frac{\pi i}{8}} 2^{\frac{1}{4}} e^{-\frac{ix^2}{2\mu}} 
e^{\frac{i}{\mu} f(x,\sqrt{2}x)} 
\left(y^{\frac{i}{2}\mu} + \mu \p_y^2 (y^{\frac{i}{2}\mu})\Bigr|_{y=\sqrt{2}x} + \Ord(\mu^2) \right)\\
 &= e^{-\frac{\pi i}{8}} 2^{\frac{1}{4} + \frac{i}{4}\mu} e^{\frac{ix^2}{2h}} 
x^{\frac{i}{2}\mu} (1+ C \mu^2 x^{-2} + \Ord(\mu^2))
\end{align*}
as $\mu$ tends to $0$ with some constant $C$. 
In fact, the second term $\mu^2 x^{-2} = \Ord(\mu)$ uniformly on $I_c(\mu)\cap \R_+$.  
Moreover we can compute 
$U_{\frac{\pi}{4}}[\phi_1^\vdash](x)$ as $-2^{-\frac{1}{2}}\mu\, U_{\frac{\pi}{4}}[ y^{-1}\phi_2^\vdash(y)](x)$ 
similarly thanks to the fact that the phase function is the same. 
Hence we obtain 
$$
U_{\frac{\pi}{4}}[\phi^\vdash](x) = 
e^{-\frac{\pi i}{8}} 2^{\frac{1}{4} + \frac{i}{4}\mu} e^{\frac{ix^2}{2h}} 
x^{\frac{i}{2}\mu} 
\begin{pmatrix}
{\displaystyle - \frac{\mu}{2 x}}\\
1
\end{pmatrix}
\left( 1 + \Ord(\mu)\right)
$$
as $\mu$ tends to $0$ uniformly on $I_c(\mu)\cap\R_+$.

On the other hand, the calculation of $U_{\frac{\pi}{4}}[\phi^\dashv](x)$ for $x \in I_c(\mu)\cap \R_-$, 
we see that the stationary point $y= \sqrt{2}x$ also lies on the finite integral path. 
Similarly we obtain   
\begin{align*}
U_{\frac{\pi}{4}}[\phi_2^\dashv](x) 
 &= e^{-\frac{\pi i}{8}} 2^{\frac{1}{4} + \frac{i}{4}\mu} e^{\frac{ix^2}{2\mu}} 
(-x)^{\frac{i}{2}\mu} \left(1+ \Ord (\mu)\right)
\end{align*}
as $\mu$ tends to $0$ uniformly on $I_c(\mu)\cap \R_-$. From the fact that 
$U_{\frac{\pi}{4}}[\phi_1^\dashv](x) = 2^{-\frac{1}{2}} \mu\, U_{\frac{\pi}{4}}[(-y)^{-1}\phi_2^\dashv(y)](x)$, we have 
\begin{align*}
U_{\frac{\pi}{4}}[\phi^\dashv](x) 
 &= e^{-\frac{\pi i}{8}} 2^{\frac{1}{4} + \frac{i}{4}\mu} e^{\frac{ix^2}{2\mu}} 
(-x)^{\frac{i}{2}\mu} 
\begin{pmatrix}
{\displaystyle - \frac{\mu}{2 x}}\\
1
\end{pmatrix}
\left(1+ \Ord (\mu)\right).
\end{align*}

\noindent
Applying the following lemma to the computation of 
$U_{\frac{\pi}{4}}[\phi^\bot](x)$ and $U_{\frac{\pi}{4}}[\phi^\top](x)$, 
we complete the proof of Proposition \ref{Asymlocal}. 
\hfill$\Box$

\begin{lemma}\label{SymNFsol}
The following relations 
holds.
$$
U_{\frac{\pi}{4}}[\phi^\bot] = \begin{pmatrix}0&1\\-1&0\end{pmatrix}{\mathcal C} U_{\frac{\pi}{4}}[\phi^\dashv],\qquad 
U_{\frac{\pi}{4}}[\phi^\top] = \begin{pmatrix}0&1\\-1&0\end{pmatrix}{\mathcal C} U_{\frac{\pi}{4}}[\phi^\vdash],
$$
where ${\mathcal C}$ is a complex conjugate operator, that is ${\mathcal C}\phi(y) = \overline{\phi(y)}$.
\end{lemma} 

\noindent
{\sl Proof of Lemma \ref{SymNFsol}}: 
Recalling the Fourier integral operators and their canonical transformations on the phase space,  
we obtain the following identities:
\begin{align*}
&{\mathcal F}_\mu^2\phi^{\vdash} = \begin{pmatrix}-1&0\\0&1\end{pmatrix}\phi^{\dashv},  
&&{\mathcal F}_\mu^2\phi^{\dashv} = \begin{pmatrix}-1&0\\0&1\end{pmatrix}\phi^{\vdash},\\ 
&U_{\frac{\pi}{4}}{\mathcal C} = {\mathcal C} U_{\frac{\pi}{4}}^{-1},
&&U_{\frac{\pi}{4}}^{-1}  {\mathcal F}_\mu^{-1} = U_{\frac{\pi}{4}},
\end{align*}
where ${\mathcal F}_\mu$ is a semi-classical Fourier transform defined by \eqref{F_mu}. 
Lemma \ref{SymNFsol} can be obtained by the following computation:
\begin{align*}
U_{\frac{\pi}{4}}[\phi^{\bot}] 
 &= U_{\frac{\pi}{4}} \begin{pmatrix}0&1\\1&0\end{pmatrix}{\mathcal C} {\mathcal F}_\mu[\phi^{\vdash}] 
 = \begin{pmatrix}0&1\\1&0\end{pmatrix}  U_{\frac{\pi}{4}} {\mathcal C} 
{\mathcal F}_\mu^{-1} {\mathcal F}_\mu^2[\phi^{\vdash}] \\
 &= \begin{pmatrix}0&1\\1&0\end{pmatrix} U_{\frac{\pi}{4}} {\mathcal C} {\mathcal F}_\mu^{-1} 
\left[\begin{pmatrix}-1&0\\0&1\end{pmatrix}\phi^{\dashv}\right]
 = \begin{pmatrix}0&1\\-1&0\end{pmatrix}{\mathcal C} U_{\frac{\pi}{4}}^{-1} {\mathcal F}_\mu^{-1}[\phi^{\dashv}] \\
 &= \begin{pmatrix}0&1\\-1&0\end{pmatrix}{\mathcal C} U_{\frac{\pi}{4}}[\phi^{\dashv}],
\end{align*}
and by the similar one concerning $U_{\frac{\pi}{4}}[\phi^{\top}]$. 
\hfill$\Box$

\section{{\sl A kind of Neumann's lemma}}\label{Neumann}

We introduce an iteration scheme for a $2\times 2$ system which reduces a function in the off-diagonal to a constant. 
\begin{lemma}\label{M_Id} 
Let $J$ be an interval on ${\mathbb R}$ and $\delta$ be a parameter. Let $\psi(z)$ be a solution of 
\begin{equation}\label{sys1}
\begin{pmatrix}
D_z + z & \delta g(z;h)\\[7pt]
\delta g(z;h) & D_z -z
\end{pmatrix} \psi(z) = 0,
\end{equation}
where the map $g: z \mapsto g(z;h)$ is $C^\infty(J;{\mathbb R})$ with $g(0;h) =1$ 
and bounded on $J$ uniformly with respect to $h \in (0, h_0]$ with all its derivatives. 
There exist $\delta_0 >0$ small enough such that for any $\delta \in (0, \delta_0]$,
we can find a $C^\infty$-matrix $M(z; \delta, h)$ given by
$$
M(z; \delta, h) = \textnormal{Id} + \sum_{k\geq 1} M_k(z;h) \delta^k,
$$ 
where $M_k(z;h)$ is bounded on $J$ uniformly with respect to $h \in (0, h_0]$
together with all its derivatives,
and then $M(z; \delta, h) \psi(z)$ is a solution of  
\begin{equation}\label{sys2}
\begin{pmatrix}
D_z + z & \delta\\[7pt]
\delta & D_z - z
\end{pmatrix} \Psi(z) =0,\quad z\in J.
\end{equation}

\end{lemma}
\begin{remark}
This kind of lemma is performed in \cite{FeGe02} and \cite{FuLaNe09_01} 
for the study of a $2\times 2$ system but depending only on one parameter. 
This lemma allows us to reduce our microlocal model to a solvable one's. 
Such a technique of reducing a symbol to a special form is well-known as a Birkhoff normal from. 
See for example \cite{Sj92_02} in the case of non-degenerate potential wells. 
And see also \cite[Section 5.4]{Co04} for the symplectic case. 
The more geometric general settings were studied in \cite{Co03}. 
\end{remark}

\noindent
{\sl Proof of Lemma \ref{M_Id}}: 
We start by constructing a $2\times 2$ matrix $M(z;\delta,h)$ such that 
$M(z;\delta,h)\psi$ satisfies \eqref{sys2} if $\psi$ is a solution of \eqref{sys1}. 
We put $\tilde g(z;h) := g(z;h) -1$ and rewrite \eqref{sys1} and \eqref{sys2} as 
\begin{align}\label{sys3}
\left(D_z {\rm Id} + z B + \delta N \right)\psi(z) &= -\delta \tilde g(z;h) N \psi(z),\\ \label{sys4}
\left(D_z {\rm Id} + z B + \delta N\right)M(z; \delta,h)\psi(z) &= 0,
\end{align}
where $B$ and $N$ are $2\times 2$ constant matrices of the forms  
$$
B = \begin{pmatrix} 1 & 0\\ 0 & -1\end{pmatrix},\qquad N = \begin{pmatrix} 0 & 1\\ 1 & 0\end{pmatrix}. 
$$
We look for a $C^\infty$-matrix $M(z;\delta,h)$ having the next form:
$$
M(z;\delta,h) = {\rm Id} + M_+(z;\delta,h),
$$
where $M_+(z;\delta,h) = \underset{k\geq 1}{\sum} \delta^k M_k(z;h)$ with
$$
M_k(z;h) = 
\begin{pmatrix}
a_k(z;h) & b_k(z;h)\\[7pt]
-\overline{b_k(z;h)} & \overline{a_k(z;h)}
\end{pmatrix}\quad \text{for}\ k\geq 1.
$$
We compute the left-hand side of \eqref{sys4} by using the system \eqref{sys3}.

\begin{align}\label{LHS}
 \left(D_z {\rm Id}  + z B +\delta N\right)&({\rm Id}+ M_+)\psi \\
 &= \left( D_z M_+ + z[B, M_+] + \delta [N, M_+] - \delta \tilde g MN \right)\psi.\nonumber
\end{align}

From the computation \eqref{LHS}, we determine $M$ satisfying 
\begin{align*}
 &\delta \left( D_z M_1 + z[B, M_1] - \tilde g N \right)\\
+\ &\sum_{k\geq 2} \delta^k \left( 
D_z M_k + z [B, M_k]  + [N, M_{k-1}] - \tilde g M_{k-1} N 
\right)= 0.
\end{align*}
Note that, concerning the matrices $B$, $N$ and $M_k$, the followings are useful. 
\begin{equation}\label{GMrelation} 
\begin{aligned}
\left[ B,\begin{pmatrix} a_k & b_k\\ -\bar{b}_k & \bar{a}_k\end{pmatrix} \right]
 &= 2\begin{pmatrix} 0 & b_k\\ \bar{b}_k & 0\end{pmatrix}, \\[7pt]
 N \begin{pmatrix} a_k & b_k\\ -\bar{b}_k & \bar{a}_k \end{pmatrix}
 &= \begin{pmatrix} -\bar{b}_k & \bar{a}_k\\ a_k & b_k\end{pmatrix},\\[7pt]
 \begin{pmatrix} a_k & b_k\\ -\bar{b}_k & \bar{a}_k\end{pmatrix} N
 &= \begin{pmatrix}  b_k & a_k\\ \bar{a}_k & -\bar{b}_k\end{pmatrix}.
\end{aligned}
\end{equation}

In the case $k=1$, one sees from \eqref{GMrelation} that the recurrence system is 
$$
-i\begin{pmatrix}a_1'(z;h)&b_1'(z;h)\\[7pt]-\overline{b_1'(z;h)}&\overline{a_1'(z;h)}\end{pmatrix}
 + 2z \begin{pmatrix}0&b_1(z;h)\\[7pt]\overline{b_1(z;h)}&0\end{pmatrix} 
= \tilde g(z;h) \begin{pmatrix}0&1\\[7pt]1&0\end{pmatrix}.
$$
We remark that two equations in the off-diagonal entries are the same each other by taking their complex conjugates 
thanks to the assumption that $g(z;h)$ is real.
From the diagonal entry, we can choose $a_1(z;h)$ as some constant independent of $h$.  
The off-diagonal entry 
\begin{equation*}\label{1-2&2-1}
\begin{aligned}
-i b_1'(z;h) + 2z b_1(z;h) = \tilde g(z;h) 
\end{aligned}
\end{equation*}
is a first order differential equation and one sees that this equation can be solved as
$$b_1(z;h) = i e^{-iz^2}\int_0^z e^{is^2}\tilde g(s;h) ds$$
with a choice of $b_1(0;h) = 0$. 
Note that $b_1(z;h) = {\mathcal O}(1)$ on $I \times (0, h_0]$ form the boundedness of $g(z;h)$.

In the case $k\geq 2$, the recurrence system is given by 
\begin{align*}
&-i\begin{pmatrix}
a_k'&b_k'\\[7pt]-\overline{b_k'}&\overline{a_k'}
\end{pmatrix} 
+ 2z 
\begin{pmatrix}0&b_k\\[7pt]\overline{b_k}&0
\end{pmatrix} 
\\
& \ + 
 \begin{pmatrix}
-( b_{k-1} + \overline{b_{k-1}}) &  \overline{a_{k-1}} - a_{k-1}\\[7pt]  
a_{k-1}-\overline{a_{k-1}} & b_{k-1} + \overline{b_{k-1}}
\end{pmatrix}
= \tilde g \begin{pmatrix}b_{k-1}&a_{k-1}\\[7pt]\overline{a_{k-1}}&-\overline{b_{k-1}}\end{pmatrix}.
\end{align*}
We also notice that each two equations in the diagonal and off-diagonal entries 
are the same each other by taking their complex conjugates.
One can solve the diagonal entry with an initial condition $a_k(0;h)=0$ as 
$$
a_k(z;h) = i\int_0^z \left( g(s;h) b_{k-1}(s;h) + \overline{b_{k-1}(s;h)} \right)ds
$$
and the off-diagonal entry with $b_k(0;h)=0$ as 
$$
b_k(z;h) = ie^{-iz^2}\int_0^z e^{is^2} \left( g(s;h) a_{k-1}(s;h) - \overline{a_{k-1}(s;h)} \right)ds.
$$
Hence we can construct recursively each entry of $M_k(z;h)$ for all $k\in{\mathbb N}$ and 
we see that $a_k(z;h)$ and $b_k(z;h)$ are ${\mathcal O}(1)$ on $J \times (0, h_0]$. 
The way to construct $a_k(z;h)$ and $b_k(z;h)$ as before and the assumption of a smoothness of $g(z;h)$ 
imply that 
$$
\| M_k(z;h) \| \leq \left( 2(||g||_{\infty} +1) \sup_{z\in J} |z| \right)^{2k-1}.
$$
Hence we set $\delta_0 = \left( 2(||g||_{\infty} +1)\, \underset{z\in J}{\sup} \, |z| \right)^{-2}$ and then for any $\delta < \delta_0$ 
the constructed matrix $M(z;\delta, h)$ is $C^\infty$ on $z$ and bounded on $J$ uniformly on $(0, h_0]$ together with its all derivatives.
\hfill $\Box$

\section{{\sl Algebraic computation}}\label{Alg_lemmas}

In order to compute the product of the transfer matrices appearing in \eqref{prodTra} , we prepare algebraic lemmas. 
First we introduce four matrices which consist of $M_2(\C)$.  
\begin{equation}\label{M2ring}
D_1 = \begin{pmatrix} 1&0\\0&0 \end{pmatrix},\quad D_2 = \begin{pmatrix} 0&0\\0&1 \end{pmatrix},\quad 
N_1 = \begin{pmatrix} 0&0\\1&0 \end{pmatrix},\quad N_2 = \begin{pmatrix} 0&1\\0&0 \end{pmatrix},
\end{equation}
whose subscript corresponds to a non-zero column. 
One sees that the set $\{ D_1, D_2, N_1, N_2\}$ is closed under the usual product $M_2(\C)$ as follows:
\begin{equation}\label{property_M}
\begin{aligned}
D_1^2 &= D_1, &D_2^2 = D_2, &&D_1D_2 = D_2D_1 = O,\\
N_1N_2 &= D_2, &N_2N_1 = D_1, &&N_1^2 = N_2^2 = O,\\
D_1N_2 &= N_2, &D_2N_1 = N_1, &&D_1N_1 = D_2N_2 = O,\\
N_1D_1 &= N_1, &N_2D_2 = N_2, &&N_1D_2 = N_2D_1 =O.
\end{aligned}
\end{equation}
We put 
\begin{align*}
T_{k,k+1} = a_k D_1 + \overline{a_k}D_2,\qquad 
T_k = b_k D_1 + \overline{b_k} D_2 + c_k N_1 + c_k N_2,  
\end{align*}
where $a_k, b_k, c_k \in \C$ and set ${\mathcal T}_k := T_k T_{k,k+1}$ for $k=0,1,2,\ldots$ with $T_0 := {\rm Id}$. 
Notice that these matrices appear in the representation of the scattering matrix.  
From the above properties, we know 
\begin{equation}\label{alg_T_k}
{\mathcal T}_k = a_kb_k D_1 + \overline{a_kb_k} D_2 + a_kc_k N_1 + \overline{a_k}c_k N_2.
\end{equation}
Thanks to these decompositions, we can understand the algebraic properties of the entries of the product of ${\mathcal T}_k$, 
which corresponds to the scattering matrix $S(\e,h)$ as in Proposition \ref{S-represent}. 
We focus on the term which includes just one factor $b_j$ or $\overline{b_j}$ in the coefficients of these matrices, 
for the reason why under our setting of this paper such term contributes the principal and sub principal terms of 
the the scattering matrix (see Lemma \ref{alg_lem1}) and the transition probability (see Lemma \ref{tau^2compu3}). 
\begin{definition}
We define the sequences $\{\sigma_n(\bm{a},\bm{c}) \}$ and $\{ \tau_{n}(\bm{a},\bm{b},\bm{c}) \}$ for $n \in \N$ 
by 
\begin{equation}\label{def_sigma}
\left\{
\begin{aligned}
\sigma_1(\bm{a},\bm{c}) &= a_0\overline{a_1} c_1,\\
\sigma_{n}(\bm{a}, \bm{c}) &= \sigma_{n-1}(\bm{a},\bm{c}) \left( {\mathcal C}^{(n)}a_{n}\right)c_{n},
\end{aligned}
\right.
\end{equation}
\begin{equation}\label{def_tau}
\left\{
\begin{aligned}
\tau_1(\bm{a},\bm{b},\bm{c}) &= a_0a_1 b_1,\\
\tau_{n}(\bm{a},\bm{b},\bm{c}) &= \tau_{n-1}(\bm{a},\bm{b},\bm{c}) \left( {\mathcal C}^{(n-1)}a_{n}\right)c_{n}
 + \sigma_{n-1}(\bm{a},\bm{c}) \left( {\mathcal C}^{(n-1)}a_{n}b_{n}\right),
\end{aligned}
\right.
\end{equation}
where the notation ${\mathcal C}^{(n)}$ stands for the $n$-th iterated composition 
${\mathcal C}\circ {\mathcal C} \circ \cdots \circ {\mathcal C}$  
with ${\mathcal C}$ the operator of taking its complex conjugate as in Lemma \ref{SymNFsol}.
\end{definition}
From the above definition \eqref{def_sigma} and \eqref{def_tau}, 
we see that $\{\sigma_n(\bm{a},\bm{c}) \}$ (resp. $\{ \tau_{n}(\bm{a},\bm{b},\bm{c}) \}$) 
is determined by $(a_0, a_1, \ldots, a_n)\in \C^{n+1}$ and $(c_1,\ldots, c_n)\in \C^n$ 
(resp. $(a_0, a_1, \ldots, a_n)\in \C^{n+1}$, $(b_1,\ldots, b_n)\in \C^n$ and $(c_1,\ldots, c_n)\in \C^n$ 
and their complex conjugates depending on the index $n$. 
Omitting the dependence on $n$ for simplicity, 
we denote these vectors by $\bm{a}\in \C^{n+1}$ and $\bm{b}, \bm{c}\in \C^n$.

\begin{lemma}\label{alg_lem1}
Let $k$ be a fixed positive integer and $a_j, b_j, c_j$ complex numbers for $j = 1, \ldots, k$ such that 
$|b_j| \ll 1$ for each $j$. We denote $\max \{ |b_1|, \dots, |b_k| \}$ by $b$. 
There exist two kinds of functions $f_k(\bm{a},\bm{b},\bm{c})$ and $g_k(\bm{a},\bm{b},\bm{c})$ satisfying $\Ord(b^2)$ 
such that 
\begin{equation}\label{alg_lemma_prod}
\begin{aligned}
\prod_{j=1}^{2k-1} {\mathcal T}_j 
 &= \left( \tau_{2k-1}(\bm{a},\bm{b},\bm{c}) + \Ord(b^3)\right)D_1
 + \left( \overline{\tau_{2k-1}(\bm{a},\bm{b},\bar{\bm{c}} )} + \Ord(b^3)\right) D_2 \\
 &\quad + \left( \overline{\sigma_{2k-1}(\bm{a},\overline{\bm{c}})} 
 + g_{2k-1}(\bm{a},\bm{b},\bm{c}) +\Ord(b^3) \right) N_1\\
 &\quad + \left( \sigma_{2k-1}(\bm{a},\bm{c})
 + f_{2k-1}(\bm{a},\bm{b},\bm{c}) + \Ord(b^3) \right)N_2,\\[7pt]
\prod_{j=1}^{2k} {\mathcal T}_j 
 &= \left( \sigma_{2k}(\bm{a},\bm{c}) + f_{2k}(\bm{a},\bm{b},\bm{c}) + \Ord(b^3) \right)D_1\\
&\ +  \left( \overline{\sigma_{2k}(\bm{a},\overline{\bm{c}})} 
 + g_{2k}(\bm{a},\bm{b},\bm{c}) +\Ord(b^3) \right) D_2\\
&\ + \left( \overline{\tau_{2k}(\bm{a},\bm{b},\overline{\bm{c}})} +\Ord(b^3) \right) N_1
 + \left( \tau_{2k}(\bm{a},\bm{b},\bm{c}) + \Ord(b^3) \right)N_2.
\end{aligned}
\end{equation}
\end{lemma}
The proof of this lemma is just algebraic computation based on the properties \eqref{property_M} and 
the definition of $\{\sigma_n(\bm{a},\bm{c}) \}$ and $\{ \tau_{n}(\bm{a},\bm{b},\bm{c}) \}$, 
so it can be omitted. 

\smallskip 
The final step for the proof of Theorem \ref{mainthm} (see  Subsection \ref{SSprodT}) requires the computation of $|\tau_n(\e,h)|^2$. 
From the definition of $\{ \sigma_{n}(\bm{a},\bm{b},\bm{c}) \}$ given by \eqref{def_sigma}, we see for any $n\in \N$, 
\begin{equation}\label{sigma_gen}
\sigma_n = \left( \prod_{l=0}^n {\mathcal C}^{(l)} a_l \right) \left( \prod_{l=1}^n c_l \right),\qquad 
|\sigma_n|^2 = \left( \prod_{l=0}^n | a_l |^2 \right) \left( \prod_{l=1}^n |c_l |^2 \right).
\end{equation}
By \eqref{sigma_gen} and from the definition of $\{ \tau_{n}(\bm{a},\bm{b},\bm{c}) \}$ (see \eqref{def_tau}), we have 
\begin{equation}\label{tau_gen}
\tau_n = \left( \prod_{l=1}^n c_l \right) \sum_{k=1}^{n} \left[ \left( \prod_{l=0}^{k-1} {\mathcal C}^{(l)} a_l \right) 
\left( \prod_{l=k-1}^{n-1} {\mathcal C}^{(l)} a_{l+1} \right) \left( {\mathcal C}^{(k-1)} b_{k} \right) \frac{1}{c_{k}} \right].
\end{equation}
In order to compute the transition probability, the form of $|\tau_n|^2$ is required and given by the following lemma:
\begin{lemma}\label{tau^2compu3}
\begin{equation}\label{tau^2compu1}
\begin{aligned}
|\tau_n|^2 &= \left( \prod_{l=0}^n |a_l|^2 \right) \left( \prod_{l=1}^n |c_l|^2 \right) \sum_{k=1}^n 
\left| \frac{b_k}{c_k} \right|^2 \\
 &\ + 2 \left( \prod_{l=1}^n |c_l|^2 \right) {\rm Re}\, \sum_{k=2}^{n} 
 \Biggl[
 \left( \prod_{l=k}^n |a_l|^2 \right) \left( {\mathcal C}^{(k)} b_k\right) \left( {\mathcal C} \frac{1}{c_k} \right)\\
 &\hspace{4cm} \sum_{j=1}^{k-1}
 \left( \prod_{l=0}^{j-1} |a_l|^2 \right)\!\! \left( \prod_{l=j-1}^{k-2} \left({\mathcal C}^{(l)} a_{l+1}^2 \right)\right) \!\!
\left( {\mathcal C}^{(j-1)} b_{j}\right) \frac{1}{c_{j}}
 \Biggr].
\end{aligned}
\end{equation}
\end{lemma}
The proof of this lemma follows directly from \eqref{tau_gen} and from the useful fact:
\begin{align*}
\left( \sum_{k=1}^n \tau_k \right) \left( \sum_{k=1}^n \overline{\tau_k} \right) 
= \sum_{k=1}^n |\tau_k|^2 + 2 {\rm Re}\, \sum_{k=2}^n \overline{\tau_k}  \sum_{j=1}^{k-1} \tau_j.
\end{align*}

\end{appendix}

\end{document}